\documentclass{jfm}
\usepackage{graphicx}
\usepackage{epstopdf, epsfig}
 \usepackage[UKenglish]{babel}
\usepackage[utf8]{inputenc}
\usepackage{adjustbox}

\usepackage{amsmath}
\usepackage{amsfonts}
\usepackage{bbm}

\newcommand{\Expec}[2]{\mathbb{E}_{#1}\left[#2\right]}
\let\oldforall\forall
\renewcommand{\forall}{\;\oldforall\,}
\newcommand{\pd}[2]{\frac{\partial #1}{\partial #2}}

\renewcommand{\vec}[1]{\underline{#1}}
\newcommand{\mat}[1]{\underline{\underline{#1}}}

\usepackage{listings}

\newcommand\pythonstyle{\lstset{
language=Python,
basicstyle=\ttm,
otherkeywords={},             
keywordstyle=\ttb\color{deepblue},
commentstyle=\ttm\color{deepgreen},
emph={rl,rlFramework},          
emphstyle=\ttb\color{deepred},    
stringstyle=\color{deepgreen},
frame=tb,                         
showstringspaces=false            %
}}

\lstnewenvironment{python}[1][]
{
\pythonstyle
\lstset{#1}
}
{}

\newcommand\pythoninline[1]{{\pythonstyle\lstinline!#1!}}

\usepackage{multirow}

\usepackage[makeroom]{cancel}

\usepackage{hyperref}
\hypersetup{
    colorlinks=true,
    linkcolor=blue,
    filecolor=magenta,      
    urlcolor=blue,
}

\usepackage{algorithm}
\usepackage{algpseudocode}
\algnewcommand{\algorithmicforeach}{\textbf{for each}}
\algdef{SE}[FOR]{ForEach}{EndForEach}[1]
  {\algorithmicforeach\ #1\ \algorithmicdo}
  {\algorithmicend\ \algorithmicforeach}
\usepackage{dirtree}

\usepackage{pdfpages}

\newcommand{\red}[1]{#1}

\graphicspath{img/}

\shorttitle{Actuator selection for flow control}
\shortauthor{R. Paris, S. Beneddine and J. Dandois}

\title{Reinforcement-learning-based actuator selection method for active flow control}

\author{Romain Paris\aff{1}
  \corresp{\email{romain.paris@onera.fr}},
  Samir Beneddine\aff{1}
 \and Julien Dandois\aff{1}}

\affiliation{\aff{1}DAAA, ONERA, Université Paris Saclay, F-92190 Meudon - France}

\begin{document}
\newcommand{\pushright}[1]{\ifmeasuring@#1\else\omit\hfill$\displaystyle#1$\fi\ignorespaces}
\newcommand{\pushleft}[1]{\ifmeasuring@#1\else\omit$\displaystyle#1$\hfill\fi\ignorespaces}
\newcommand*{\MyPath}{.}
\newcommand*{\img}{.}

\maketitle

\begin{abstract}
This paper addresses the issue of actuator selection for active flow control by proposing a novel method built on top of a reinforcement learning agent. Starting from a pre-trained agent using numerous actuators, the algorithm estimates the impact of a potential actuator removal on the value function, indicating the agent's performance. It is applied to two test cases, the one-dimensional Kuramoto-Sivashinsky equation and a laminar bi-dimensional flow around an airfoil at Re=1000 for different angles of attack ranging from 12 to 20 degrees, to demonstrate its capabilities and limits. \red{The proposed actuator-sparsification method relies on a sequential elimination of the least relevant action components, starting from a fully developed layout. The relevancy of each component is evaluated using metrics based on the value function.} Results show that, while still being limited by this intrinsic elimination paradigm \red{(\textit{i.e.} the sequential elimination)}, actuator patterns and obtained policies demonstrate relevant performances and allow to draw an accurate approximation of the Pareto front of performances versus actuator budget. 
\end{abstract} 


\section{Introduction}


Flow control is one of the most used methods to improve aero/hydrodynamic qualities of vehicles. It encompasses a spectrum of methods from passive control to closed-loop non-linear control. Passive control through shape optimisation and/or use of riblets, vortex generators, etc. has long been a go-to strategy thanks to its simplicity and robustness \citep{BruneauMortazavi2008,SeshagiriCooperTraub2009,JoubertLePapeHeineEtAl2013,EvansHamedGorumluEtAl2018}. 
Yet, better performance may often be reached through active control. The most simple approaches are open-loop strategies where the control law is predetermined irrespective of the state and response of the flow. While demonstrating fitness to reach control goals in numerous cases \citep{SeifertBacharKossEtAl1993,SeifertPack1999}, their general energetic cost and inability to adapt to the environment response and correct control error pledge for closed-loop methods.

Closed-loop control design is nowadays a well-documented and studied domain. Most synthesis methods rely on a linear model of the controlled system to optimise a feedback control law under criteria of stability, robustness, tracking error minimisation and/or disturbance rejection. The underlying linear assumption is the main challenge stated by the transposition of such design strategies to flow control. In particular, the system dimensionality, non-linear response, and simulation cost are issues specific to fluid mechanics. A wide literature proposed adapted methods to design closed-loop control laws with these constraints, as highlighted by \citet{BruntonNoack2015}. Among the recent related work, one may cite \citet{SeidelFagleyMcLaughlin2018}, who develop a multi-step method to design controllers, using open-loop responses to forcings and building a reduced-order model of the system.

Flow control aims to mitigate or suppress the detrimental effects of some dynamical mechanisms. Most of these well-documented issues concern noise emission, flow separation \citep{ZamanMcKinzieRumsey1989,GuptaAnsell2019} causing a drop in performances (increasing drag and/or decreasing lift) or alternated loads inducing critical structural fatigue. Airfoil flow separation control \citep{WuLuDennyEtAl1998} has been a matter of interest for already a few decades, with studies using a wide variety of control methods at various flow and stall regimes \citep{SeifertDarabiWyganski1996,AmitayGlezer2002,ShimomuraOgawaSekimotoEtAl2017,YehTaira2019}.

In the last few years, machine learning, and deep learning in particular, has demonstrated remarkable performances in a wide variety of fields such as natural language processing \citep{OtterMedinaKalita2020}, image recognition \citep{GhoshDasDasEtAl2019} or robotics \citep{IbarzTanFinnEtAl2021}. This progress is mainly due to the rise in accessible computing power and the use of neural networks (NN) that act as quasi-universal function approximators that can be optimised fairly easily. This explains the flourishing literature investigating the potential of such techniques in fluid mechanics, as highlighted by \citet{Brunton2020,VinuesaBrunton2021}.

Mirroring computer vision and image classification methods, multiples studies such as \citet{ZhangSungMavris2018,HuiBaiWangEtAl2020,BhatnagarAfsharPanEtAl2019,SekarZhangShuEtAl2019} aim at deriving aerodynamic coefficients, predicting flow fields or airfoil shapes using convolutional neural networks (CNN) in a supervised manner. These studies take advantage of the NN's interpolation ability, optimising the bias-variance trade-off and surpassing the performances of most response-surface-based methods such as kriging. This enables faster design cycles, relying on these intermediate fidelity models. Linear decomposition methods such as Proper Orthogonal Decomposition (POD) or Dynamic Mode Decomposition (DMD) and their variations are increasingly challenged by non-linear reduced order modelling tools that demonstrate their capacity to compress state information further without losing in precision \citep{PichiBallarinRozzaEtAl2021,FukamiHasegawaNakamuraEtAl2020,KneerSayadiSippEtAl2021,WangXiaoFangEtAl2018,LeeCarlberg2020}. Some studies, such as \citet{LuschKutzBrunton2018}, even integrate a reduced order dynamical model, where pre-defined constraints (such as linearity and sparsity) help understand full-state dynamics. Recurrent structures such as long short-term memory cells (LSTM) provide a convenient way to integrate model dynamics \citep{MohanGaitonde2018,HasegawaFukamiMurataEtAl2020a}. Others leverage generative capacities of generative adversarial networks (GAN) and (variational) auto-encoders (VAE) to generate boundary layer velocity profiles, realistic homogeneous turbulence for direct numerical simulations (DNS) \citep{MilanoKoumoutsakos2002,KochkovSmithAlievaEtAl2021} or conversely to model turbulence \citep{DuraisamyIaccarinoXiao2019}. Yet these neural structures display poor extrapolation capabilities. Some authors try to mitigate this issue by embedding some key characteristics of fluid mechanics such as mass conservation or invariances in the models \citep{DjeumouNearyGoubaultEtAl2021,WongOoiGuptaEtAl2021,WangWuXiao2017}.




Flow control also follows this trend of increasing integration of advanced machine learning methods, notably leveraging the paradigm of reinforcement learning (RL), based on the idea of trial-and-error. RL consists in evaluating a given control law (also called a policy) on a target system (referred to as the environment). The evaluation data is then used to tweak the control law to maximise a given performance metric, with the underlying idea of promoting beneficial actions or strategies with respect to the metric. In the case of deep reinforcement learning, the policy is embodied by a neural network structure tasked with providing a control action given observed measurement. Exploring new control strategies plays a decisive role in the performance of such methods. \red{A flourishing literature \citep{GarnierViqueratRabaultEtAl2019,RabaultRenZhangEtAl2020,VonaLauga2021,LiZhang2022,MaoZhongYin2022,ZhangWangWangEtAl2021}, implements these methods and demonstrates their significant potential for flow control.} \red{Applications range from the control of a Rayleigh-Bénard convection cell \citep{BeintemaCorbettaBiferaleEtAl2020} to destabilised 1D liquid films \citep{BelusRabaultViqueratEtAl2019} and even to transitional flows, that are of crucial interest for the control of turbulence, for instance with \citet{RenRabaultTang2021} who successfully implement RL-trained flow control on a cylinder wake.}

DRL was implemented on a cylinder flow by \citet{KoizumiTsutsumiShima2018}, using a \textit{Q-learning} algorithm to learn how to stabilise the cylinder wake efficiently. Q-learning is one of the two main families of DRL methods, which consists in implicitly encoding the policy in a Q-network that provides the value of every state-action pair, the policy being the action that maximises this value for each given state. \citet{BucciSemeraroAllauzenEtAl2019} used Deep Deterministic Policy Gradient (DDPG) to steer the Kuramoto-Sivashinksy equation to its fixed points, and \citet{ShimomuraSekimotoOyamaEtAl2020} used Deep Q-Networks (DQN) to optimise the frequency of plasma actuator bursts to control airfoil flow separation. These methods make use of a self-consistency equation for their optimisation and thus display a well-known tendency to be unstable. Most of \textit{policy optimisation} algorithms, the other main family of DRL training methods, derive from Proximal Policy Optimisation (PPO) introduced by \citet{SchulmanWolskiDhariwalEtAl2017}. This algorithm directly implements the idea of promoting successful control action and avoiding detrimental ones. PPO was used by \citet{RabaultKuchtaJensenEtAl2019} to control a low Reynolds number cylinder wake using surface-mounted jets. \citet{WangMeiAubryEtAl2022} also resorted to PPO to control a low-Reynolds confined bi-dimensional airfoil flow using three suction-side synthetic jets to reduce drag. In this paper, a variant named Proximal Policy Optimisation with Covariance Matrix Adaptation (PPO-CMA) \citep{HaemaelaeinenBabadiMaEtAl2018} is used as the base learning algorithm.


Fluid mechanics contrasts with classical artificial intelligence application domains by its sample cost, dimensionality, and non-linearity, as mentioned earlier. These factors represent challenges that are characteristic of the difficulties encountered in the scale-up of most machine learning methods to real-world applications. Experimental application is an intermediate step between simple numerically simulated cases and pre-industrial prototypes. Yet, for such data greedy methods and for specific cases, experiments may be less expensive per sample than numerical simulation since experiments tend to have high fixed costs and lower marginal costs, contrary to numerical simulation. While, for numerical simulation, state measurements and forcing actions are only limited by the stability of numerical schemes, experiments require sensors to be non-intrusive and actuators to be constrained to both physical integration restrictions and intrinsic performances, for instance in terms of frequency bandwidth or power. 

In this context, the issue of sensor and actuator location or selection becomes critical. Having an efficient and parsimonious control setup motivates the emergence of all the following methods. In their multi-step heuristic approach to closed-loop flow control design, \citet{SeidelFagleyMcLaughlin2018} propose to rely on the correlation between observed relevant phenomena and sensor signal to choose observations and on clues provided by POD mode decomposition of the flow instabilities to choose actuators location. \citet{Cohen2006} and \citet{Willcox2006} also leverage POD analyses to derive optimised sensor placement.


Multiple studies rely on adjoint sensitivity analysis \citep{Chomaz2005} and on the "wavemaker", the overlap between the direct and adjoint sensitivity modes, introduced by \citet{GiannettiLuchini2007} to derive appropriate sensor and actuator placement. \citet{LiZhang2022} used a computation of the wavemaker on a confined cylinder flow to lay their sensors used as feedback for a reinforcement-learning-trained policy. \citet{NatarajanFreundBodony2016} resorted to the wavemaker to optimally locate both sensors and actuators in a diffuser flow.

\citet{SashittalBodony2021} applied a related method on a data-driven, linearised model of their systems to position their sensors. They used this method to control both the linearised complex Ginzburg-Landau equation and a flow around an inclined flat plate. Linear-quadratic-Gaussian control on balanced truncated reduced models is employed to derive optimal sensor and actuator placement using gradient descent methods \citep{Manohar2018,YaoSunHemati2020}. Modelling the control system as a linear plant is also used by \citet{BhattacharjeeHematiKloseEtAl2018} who take advantage of the eigensystem realisation algorithm (ERA) to compare the controllability (in a $\mathcal{H}_2$ framework) of multiple jet actuators laid on the suction of an airfoil to select the best one depending on the performance criterion (lift or angle of attack upon flow separation).

\citet{Oehler2018} and \citet{JinIllingworthSandberg2022} used both optimal estimation and full-state information control to derive optimal sensor and actuator placement for the complex linearized Ginzburg-Landau system and a low Reynolds number cylinder wake, respectively. \citet{YehTaira2019} also employed resolvent-based analyses to discover the optimal forcing variable and location for an actuator aiming at preventing airfoil flow separation. \citet{LuharSharmaMcKeon2014} leverage resolvent analyses to assess the potential of opposition control for drag reduction and turbulence control.

Among non-linear actuator methods, one may cite the study of \citet{Rogers2000}, who derived a set of actuator layouts on a stealth bomber to satisfy manoeuvrability goals using a genetic algorithm. \citet{ParisBeneddineDandois2021} proposed a sensor selection algorithm using stochastic gates and leveraging the RL paradigm to filter out sensor measurement while preserving performance as much as possible.

Most of the previously cited actuator location/selection methods rely on linear analyses. In the current study, we aim at introducing a reinforcement-learning-based method, leveraging some ideas previously introduced in \citet{ParisBeneddineDandois2021} but this time to select actuators instead of sensors.
The present paper provides an in-depth study of a new sparsity-seeking algorithm for actuators, a question that has never been addressed in the context of flow control by RL to our knowledge. By selecting two test cases, each one being challenging in their own way, the paper gives a critical analysis of the algorithm. This allows a more general discussion about the difficulties encountered when one tries to use RL for automatic actuator placement. Therefore, the goal of the present paper is twofold: introducing the first RL algorithm that tackles the essential question of actuator placement for flow control, and identifying future challenges and research directions to pave the way for upcoming work on the topic.
 Before describing the proposed action-sparsity-seeking algorithm, the reinforcement learning method used as a base to the algorithm is introduced in part \ref{sec:ASPPOCMA}. Both test cases are then described and results of the method are discussed in sections \ref{sec:KS} and \ref{sec:NACA}. At last, a discussion about the limitations of the proposed sparsity-promoting method is led.

\section{An action-sparsity-seeking algorithm}\label{sec:ASPPOCMA}
\renewcommand*{\MyPath}{P1_ASPPOCMA}
\renewcommand*{\img}{P1_ASPPOCMA/img}
\subsection{The generic reinforcement learning loop}
\begin{figure}
    \centering
    \includegraphics[width=0.5\textwidth,page=3]{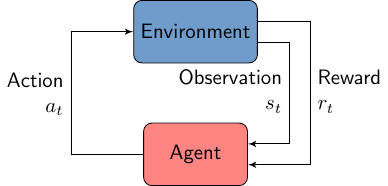}
    \caption{The reinforcement learning feedback loop}
    \label{fig:RL_loop}
\end{figure}

This study relies on the derivation of closed-loop control laws (called policies) trained by reinforcement learning. As described by figure \ref{fig:RL_loop}, the environment encloses the experimental setup or the numerical simulation and provides partial state observations $s_t$ and a reward $r_t$, quantifying the instantaneous fitness of the current state with respect to a predefined performance metric. The environment can receive forcing actions $a_t$ that modify its behaviour when marched forward in time: $s_{t+1}\sim T(\cdot|s_t,a_t)$. The agent has two roles. It provides control actions $a_t$ based on observations $s_t$ following its policy $\pi$: $a_t\sim\pi(\cdot,s_t)$.
It is also tasked with implementing the training algorithm, that uses the collected data samples $(s_t,a_t,r_t)$ to tune the parameters of policy $\pi$ in order to maximise the expected return $R_t=\Expec{s_t\sim T,a_t\sim\pi}{\sum_{t=0}^{\infty}\gamma^tr_t}$ for every partial state $s_t$, where $\gamma\in[0,1]$ is an actualisation scalar parameter. It quantifies the preference for immediate rewards rather than future rewards, and is typically set to values slightly below 1.

\subsection{Proximal Policy Optimisation with Covariance Matrix Adaptation} 

The reinforcement learning algorithm used here is \textit{on-policy}. This means that, at anytime in the training, the data used for optimisation has been provided by the previous control episode (also called roll-out) using the current policy $\pi$ only. Thus, the training can be split into epochs, made of one or several episodes (data collection phase, evaluation of the current policy) and an optimisation phase using the data collected.

Proximal Policy Optimisation (PPO) is one of the state-of-the art on-policy approaches proposed by \citet{SchulmanWolskiDhariwalEtAl2017}. It relies on a actor-critic structure, where the agent contains two neural networks: an actor ($\pi$) and a critic ($V$). The actor provides the control action $a_t$ using $s_t$ as input. This action is sampled stochastically from a normal distribution $a_t\sim\pi_\theta(\cdot|s_t) = \mathcal{N}\left(\mu_\theta(s_t),\sigma\right)$, where $\theta$ are the trainable parameters (weights and biases) of the actor neural network, $\mu_\theta$ the deterministic output of $\pi$ (the optimal action to implement according to the actor) and $\sigma$ a predefined standard deviation. This random sampling of the control action helps solving the exploration-exploitation dilemma efficiently for high-dimensional and continuous state-action spaces. Thus, this strategy enables to explore only the vicinity of the deterministic (and best) trajectory in the state-action space, and $\sigma$ is simply set to zero during policy testing. The critic is trained to output an estimate $V_\phi(s_t)$ of the value $V^\pi(s_t)$ of the current observed partial state, where $\phi$ are the parameters of the critic neural network. This value is computed as the expected return $R_t=\Expec{s_\tau\sim T,a_\tau\sim\pi}{\sum_{\tau=t}^{\infty}\gamma^\tau r_\tau}$ previously introduced, the expected actualised sum of the reward under the current policy $\pi$. $V_\phi$ is trained in a supervised fashion using the observed returns of the current episode as target. $\pi$ is trained to maximise the likelihood of successful actions (the ones that provide an increased return) and to reject poorly valued ones. More details can be found in the original article \citep{SchulmanWolskiDhariwalEtAl2017}.

\begin{figure}
    \centering
    \includegraphics[page=6,width=0.5\textwidth]{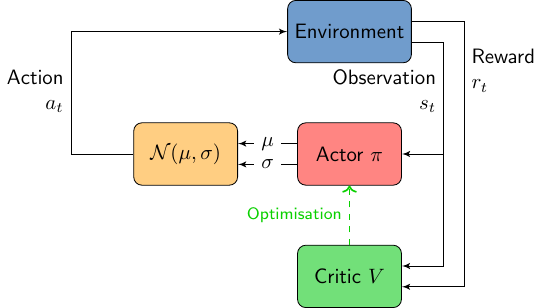}
    \caption{PPO-CMA agent structure: compared to PPO, the actor has one extra output $\sigma$ which allows for a dynamic adaptation of the exploration.}
    \label{fig:PPOCMA_structure}
\end{figure}

The standard PPO algorithm uses a clipping mechanism to prevent excessively large policy updates, which ``can prematurely shrink the exploration variance" according to \citet{HaemaelaeinenBabadiMaEtAl2018}. To address this issue, these authors introduced Proximal Policy Optimisation with Covariance Matrix Adaptation (PPO-CMA), as a variant of PPO. The main difference with the standard version of PPO is that $\sigma$ becomes a trained output of $\pi$ alongside with $\mu_\theta$. The training of $\sigma$ is performed using bufferized data of the last episodes, which is a slight diversion from the on-policy paradigm. This enables a dynamic exploration, large at the beginning of training, smaller at convergence. In this study, the covariance matrix $\sigma$ is diagonal, thus $\sigma$ is cast to a vector of length $n_{act}$ in the following. The structure of PPO-CMA is summarised by figure \ref{fig:PPOCMA_structure}. Neural networks' architecture and most of the hyper-parameters of the learning algorithm are summed-up in table \ref{tab:params} and fixed (unless explicitly stated). Their values have been set by trial-and-error, the optimality of the hyper-parameters choice is not guaranteed and is out of the scope of the current study.

\subsection{A stochastic gating mechanism to sparsify actions}
\begin{figure}
    \centerline{
    \includegraphics[page=22,width=0.5\textwidth]{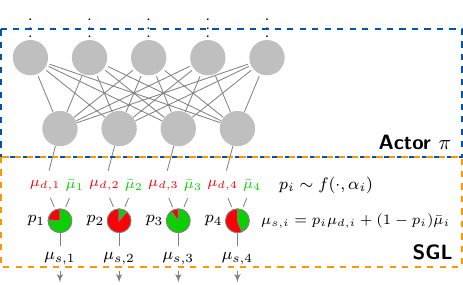}
    }
    \caption{Structure of the Stochastic Gated Layer (SGL) used to filter the actor output. Here $\mu_d$ and $\mu_s$ respectively stand for $\mu_{dense}$ and $\mu_{sparse}$.}
    \label{fig:SGL_layer}
\end{figure}

In this part, the proposed action-sparsity-seeking mechanism is introduced. The proposed structure aims at learning a binary mask $\vec{p}\in\{0,1\}^{n_{act}}$, reducing the number of non-null actions components to a prescribed amount. Let $\vec{a}_{dense} \sim \pi = \mathcal{N}(\vec{\mu}_{dense},\vec{\sigma}_{dense})$ be the stochastic action provided by the actor. As illustrated by figure \ref{fig:SGL_layer}, the gated action $\vec{a}_{sparse}$ is defined so that:
\begin{align*}
    \vec{a}_{sparse} =\vec{p}\odot\vec{a}_{dense}+(1-\vec{p})\odot\vec{\bar{a}},
\end{align*}
with $\odot$ the scalar dot product, and $\vec{\bar{a}}$ being substitution values. In our case $\vec{\bar{a}}$ is a null vector (the action is simply clipped), thus the previous relation reduces to $$\vec{a}_{sparse}=\vec{p}\odot\vec{a}_{dense}.$$ When $p_i=0$, the i$^{\text{th}}$ component $a_{dense,i}$ is always filtered through the gate layer ($a_{sparse,i}=0$). Conversely, when $p_i=1$, the probability to filter out the action is null. During the sparsity-seeking phase of training, the mask $\vec{p}$ can take continuous values in $[0;1]^{n_{act}}$. At convergence and for evaluation, this learned mask $\vec{p}$ is bound to be constant and contains only binary values. As values of $\vec{\mu}_{dense}$ and $\vec{\sigma}_{dense}$ need to be provided for the optimisation of the actor, clipping is also applied on these vectors, by ``replaying" the same values of $\vec{p}$ as for the sparse $\vec{a}_{sparse}$ in the following way:
\begin{align*}
    \vec{\mu}_{sparse} &=\vec{p}\odot\vec{\mu}_{dense}\\
    \log\vec{\sigma}_{sparse} &=\vec{p}\odot\log\vec{\sigma}_{dense}+(1-\vec{p})\left(-\frac{1}{2}\log(2\pi)\right).
\end{align*}
This clipping of $\log\vec{\sigma}$ is defined to keep the log-likelihood of an action independent of the number of clipped components (for which the value is deterministically set to $0$).

During training, the value of $\vec{p}$ is either sampled on Bernoulli distribution or ``replayed" for optimisation using a structure similar to the Stochastic Gating Layer (SGL) proposed by \citet{Louizos2017} and has been previously used by \citet{ParisBeneddineDandois2021} to reduce the number of observations for flow control by RL. It is used here in a simplified version, $\vec{p}$ is derived \red{as a vector of Bernoulli trails}:
\begin{equation}
    p_i= f(\alpha_i,u_i) =  \begin{cases}
    1, & \text{if } u_i\leq \alpha_i\\
    0,              & \text{otherwise}
\end{cases}
\end{equation}
where $\vec{u}\in[0,1]^{n_{act}}$ is a random vector and $\vec{\alpha}$ is a trainable vector that sets the probability for the gate to be open. The sampling of $\vec{u}$ depends on the training phase as summarised in table \ref{tab:SGL_train_table}. During the sampling phase, the values $\vec{u}_{samp}$ are sampled from a uniform distribution and stored in the training buffer $\mathcal{B}$ alongside observations, actions and rewards. These values are then used during the optimisation phase to replay the state $\vec{p}$ of the gate and back-propagate faithful gradients.

\begin{table}
    \centering
    \begin{tabular}{cccc}
        & Sampling & Optimisation & Evaluation\\
        $\vec{u}$ & $\vec{u}_{samp}\sim \mathcal{U}^{n_{act}}(0,1)$ & $\vec{u}_{opt} = \vec{u}_{samp}$ & $\vec{u}_{eval} = 0.5$\\
        $\vec{p}$ & $\mathbbm{1}_{\vec{u}_{samp}<\vec{\alpha}}$ & $\mathbbm{1}_{\vec{u}_{samp}<\vec{\alpha}}$ & $\mathbbm{1}_{0.5<\vec{\alpha}}$\\
        $P(p_i = 1)$ & $\alpha_i$ & $\alpha_i$ & $\mathbbm{1}_{0.5<\alpha_i}$
    \end{tabular}
    \caption{Values of $\vec{u}$ and $\vec{p}$ during the different phases of the training process}
    \label{tab:SGL_train_table}
\end{table}

\subsection{Sequential elimination process and ranking metric}

\begin{figure}
    \centering
    \raisebox{-0.5\height}{ \includegraphics[width=.58\textwidth,page=51]{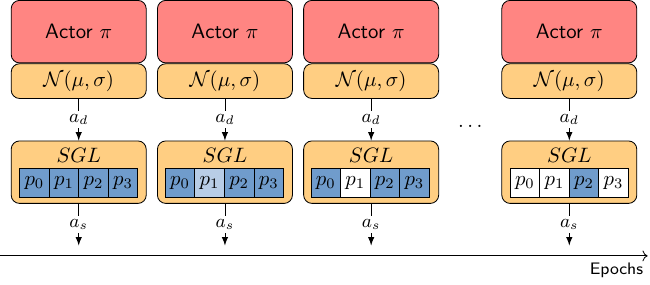}}
    \raisebox{-0.5\height}{ \includegraphics[width=.39\textwidth,page=53]{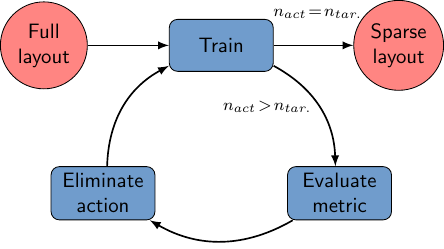}}
    \caption{(left) Schematics of the proposed elimination process. Gates' opening probability are represented as blue squares in the SGL layer, their colour saturation indicates the opening probability. (right) Illustration of the sequential elimination process proposed. The algorithm iterates over this loop until the breaking condition (i.e. the required number of eliminated actuators) is met.}
    \label{fig:seq_elim}
\end{figure}

The actor-critic structure is first trained in a standard fashion, with $\alpha_i=1$ values guaranteeing fully open gates. Once this first phase is completed and the desired control performance reached, the sparsity-seeking phase starts. As shown by figure \ref{fig:seq_elim} (left), the main goal is to close a prescribed number of gates by tuning $\vec{\alpha}$. The update of $\vec{\alpha}$ is performed one component at a time, in a sequential fashion, gradually tuning $\alpha_i$ from 0 to 1 corresponding to the elimination of the $i^{th}$ action component. The choice of the eliminated action is performed based on a Polyak averaged metric (denoted as $m$) computed on each active component, the worst ranked component being chosen to be eliminated next. Figure \ref{fig:seq_elim} illustrates this sequential elimination process, that iterates over a three-phase loop:
\begin{enumerate}
    \item A first standard training phase where the policy is trained with all its active action components,
    \item A second phase where the ranking metric $m$ is estimated,
    \item A third phase consisting into the action clipping itself.
\end{enumerate}


The algorithms keeps on iterating this loop until the prescribed number of active components is met, as illustrated by figure \ref{fig:seq_elim} (right). In the present the metric ranking is based on a ``what-if" analysis, where the effects of clipping an action component on the overall performance are estimated, by computing the expected clipped value function $V^i$ over the system's dynamics:
\begin{align*}
    m_i = \Expec{s\sim T,a\sim\pi^i}{V^i(s)}\;\forall\,i\in\{0,\dots,n_{act}\},
\end{align*}
where $\pi^i = (1-\delta_{i,j})\pi_j$, the $i^{th}$-component clipped actor. Once the component to be eliminated is chosen, the corresponding component $\alpha_i$ is tuned down slowly on a predefined number of training runs, to let the agent adapt to the elimination before fine-tuning the agent on the new action layout.

\subsection{Practical implementation}\label{sec:2B_pract_impl}
To properly learn the metric $m$ in phase (ii) of the elimination loop, roll-outs have to be performed on the environment using $\pi^i$ in order to collect relevant observation and return values. This clipped data is buffered (in a buffer $\mathcal{B}_i$). Similarly to the original actor-critic neural structure of the PPO-CMA, $V^i$ functions are embodied by neural networks (one extra neural network per action component). During that metric-evaluation phase, these networks are trained using the specific buffered data collected on clipped roll-outs, in a identical supervised fashion as for the standard value function $V$ using:
\begin{align*}
    \mathcal{L}_{V^i} = \frac{1}{|\mathcal{B}_i|}\sum_{(s_t,R_t)\in\mathcal{B}_i}\left(R_t-V^i(s_t)\right)^2
\end{align*}
as training loss. Note that these extra roll-outs represent a significant cost overhead. This may strongly impact training time when running on costly environments and if the number of action components $n_{act}$ is high. To save on computation time, clipped roll-outs corresponding to already eliminated components are not performed and their corresponding value function network is no longer trained. This identified limitation (extra roll-outs and networks) of the present algorithm is discussed in the conclusion. 

\begin{figure}
    \centering
    \begin{minipage}{0.49\textwidth}
        \includegraphics[page=25,width=\textwidth]{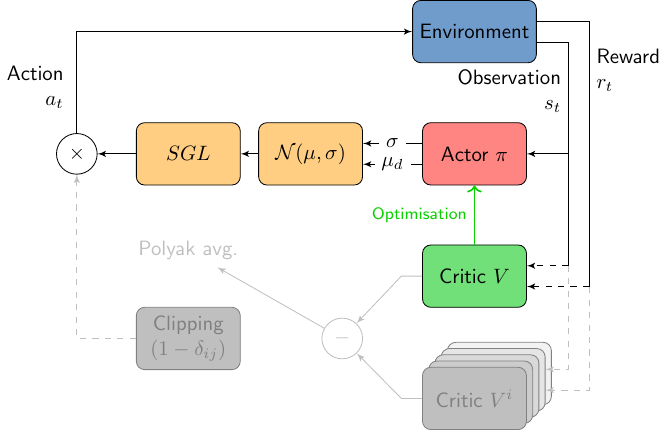}
        \vspace{0.2em}\\
        \includegraphics[page=27,width=\textwidth]{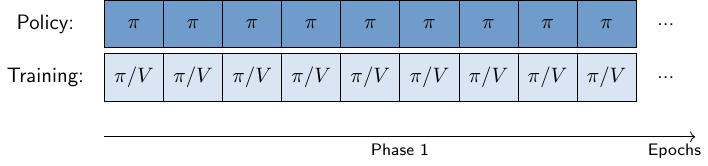}
    \end{minipage}
    \begin{minipage}{0.49\textwidth}
        \includegraphics[page=26,width=\textwidth]{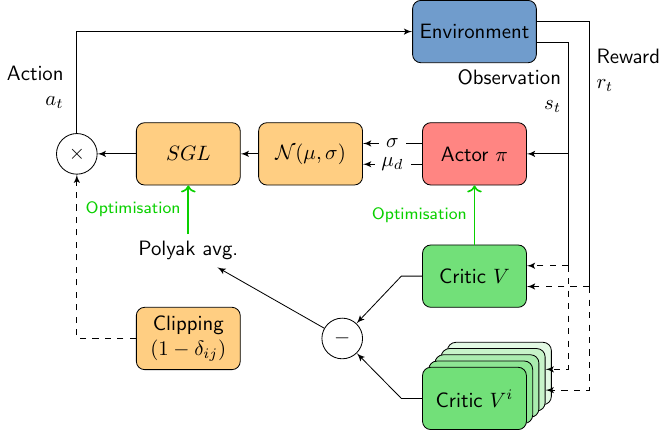}
        \vspace{0.2em}\\
        \includegraphics[page=28,width=\textwidth]{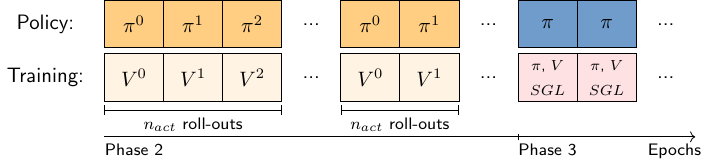}
    \end{minipage}
    \caption{AS-PPO-CMA agent structure and training operations. (left) First training phase. Greyed-out structures are not used. (right) Second and third training phase. Phases 2 and 3 see the alternance of policy $\pi$ and main critic $V$ training with specific training for each critic $V^i$.}
    \label{fig:ASPPOCMA_structure}
\end{figure}

\begin{figure}
    \centering
    \includegraphics[width=.9\textwidth]{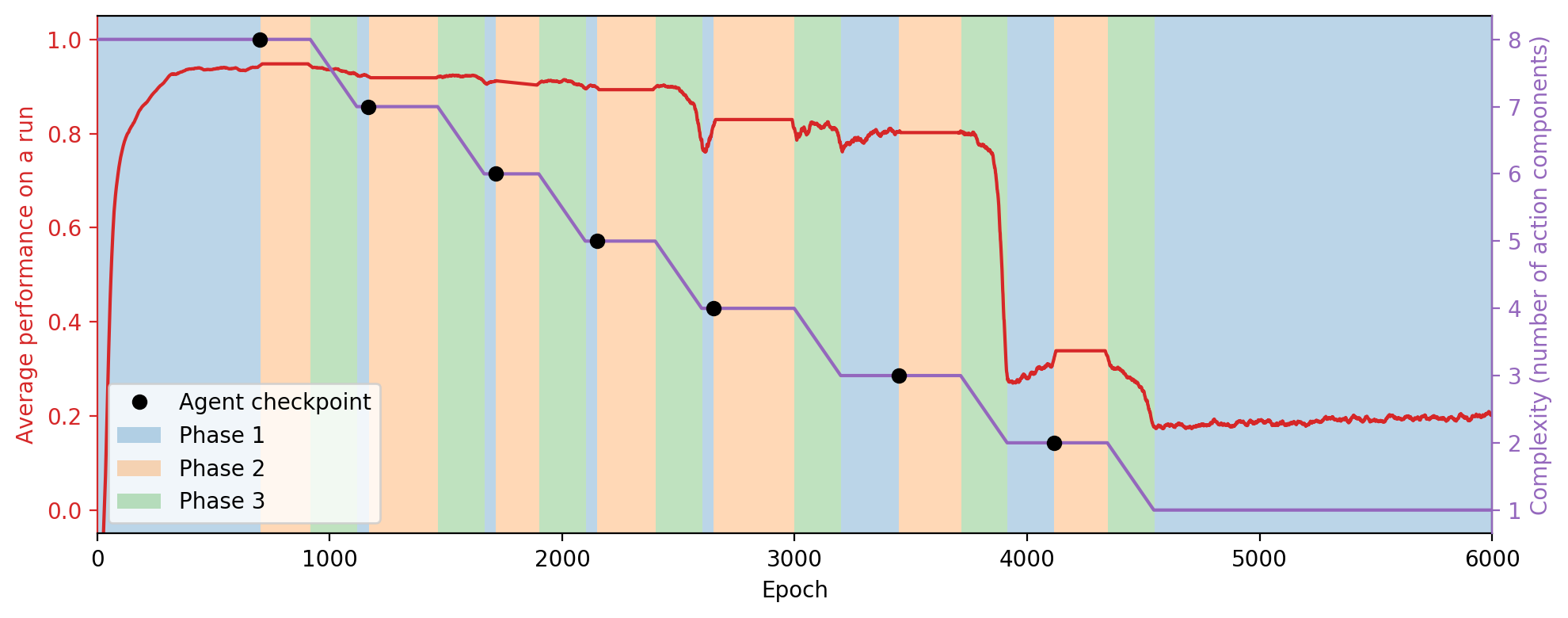}
    \caption{Example of the one-by-one elimination strategy drawn from a Kuramoto-Savishinsky test case (section \ref{sec:KS}). The performance curve (red line) has been smoothed using rolling average of length 20 for the sake of readability. Shaded areas in the background illustrate learning phases. Black dots represent the evaluation runs and the checkpoint back-ups of the agent performed during training.}
    \label{fig:ASPPOCMA_one-by-one}
\end{figure}

The practical implementation of the three phases of the loop is described by figure \ref{fig:ASPPOCMA_structure}, with phase (i) described on the left-hand side, corresponding to a standard training, with SGL $\alpha$ values kept frozen and phases (ii) and (iii) being reported on the right-hand side, where clipped value functions are trained on collected data, then the elimination action component is clipped. In practice, proceeding from phase (i) to phase (ii) and from (ii) to (iii) is only done if pre-defined performance stability criteria are met. Otherwise, the agent remains in the current phase until it complies to these. This allows for flexible elimination scheduling, saving training epochs. These stability criteria are based on the comparison of different Polyak averages of the metric $m$ and on the standard value function (depending on the phase). Figure \ref{fig:ASPPOCMA_one-by-one} provides an illustrative example of the elimination process. During phase (iii) of the process, the $\alpha_i$ component corresponding to the eliminated actuator is linearly tuned down from $1$ to $0$ (over 200 epochs for both of the test-cases later introduced).

\renewcommand*{\MyPath}{P2_KS}
\renewcommand*{\img}{P2_KS/img}

\section{The one-dimensional Kuramoto-Sivashinsky equation}\label{sec:KS}
The first test case considered is the control of the one-dimensional Kuramoto-Sivashinsky  (KS) equation. As shown later in this section, this case is very interesting because it is computationally inexpensive, allowing to perform a brute-force search of the optimal actuator layouts, yet complex when analysing the obtained optimal layouts. It is therefore a particularly challenging test case for the present algorithm that highlights some of its strengths and weaknesses.

\begin{figure}
    \centerline{
    \includegraphics[height=0.27\textwidth]{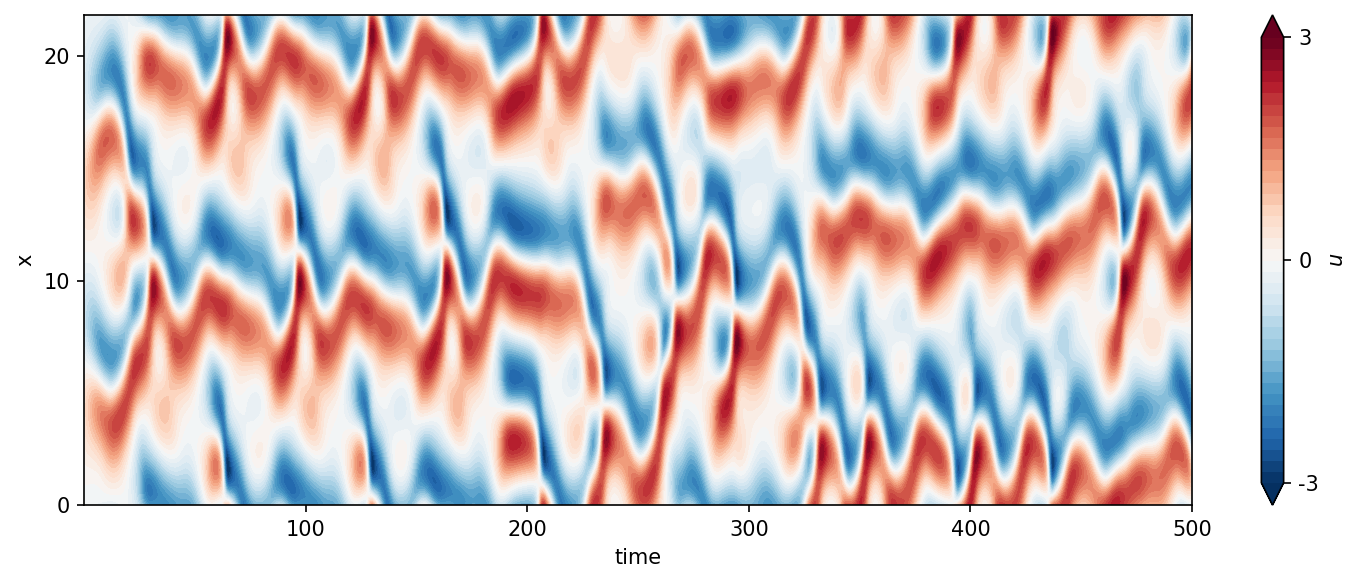}
    \includegraphics[height=0.27\textwidth]{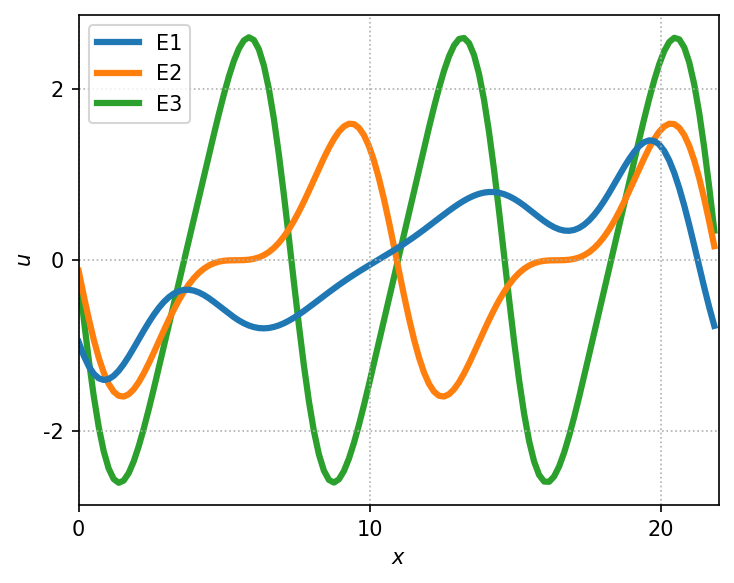}
    }
    \caption{(left) Spatio-temporal representation of the dynamics of the KS equation on $500$ non-dimensional time units, corresponding to $2000$ control steps. (right) Shape of the three fixed points of the KS equation.}
    \label{fig:KS_0}
\end{figure}

The KS equation is a well-studied fourth-order partial differential equation exhibiting a chaotic behaviour and describing the unstable evolution of flame fronts \citep{Sivashinsky1980}. On a periodic domain of length $L$, the KS equation reads:
\begin{align}
    \pd{u}{t}+u\pd{u}{x}+\frac{\partial^2 u}{\partial x^2}+\frac{\partial^4 u}{\partial x^4} = a,  \label{eq:KSequ}\\  \forall t: u(0,t) = u(L,t),\nonumber
\end{align}
where $a$ is the control action forcing later described. For small spatial domains (here $L=22$), the KS equation exhibits three fixed points (i.e. three steady solutions of (\ref{eq:KSequ}) named E1, E2 and E3) and low-dimensional instabilities similarly to some low-Reynolds number Navier-Stokes flows, as illustrated by figure \ref{fig:KS_0}. \red{Further mathematical details about the control of the KS equation can be found in the work of \citep{BucciSemeraroAllauzenEtAl2022} .}

The numerical simulation is carried out using a 64-mode Fourier spatial decomposition and a third-order semi-implicit Runge-Kutta scheme (implicit formulation for the linear terms, explicit for the non-linear term) marched in time with a time-step of $0.05$. This numerical setup is based on the work of \citet{BucciSemeraroAllauzenEtAl2019} and a code from \href{https://github.com/jswhit/pyks/blob/master/KS.py}{pyKS}. The control term is also designed to mimic spatially localized Gaussian forcing actions:
\begin{align*}
    a(x,t) = \sum_{i=0}^{n-1}a_i(t)\frac{1}{\sqrt{2\pi\sigma}}\exp{\left(-\frac{(x-x^{act}_i)^2}{2\sigma^2}\right)},
\end{align*}
where $n$ is the number of control actions, $x^{act}_i\;i\in \{0,...,n-1\}$ the locations of the centers of Gaussian kernels and $a_i$ the amplitude of each forcing implemented around $x_i$. Unless otherwise stated, the forcing action has $8$ forcing components implemented at equi-spaced locations ($x^{act}_i\in\{0,1,...,7\}L/8$), $a_i\in[-0.5,0.5]$ and $\sigma=0.4$. The partial state observations are provided by measurements of $u$ interspersed between control action locations so that $x^{obs}_i\in\{1,3,5,7,9,11,13,15\}L/16$.

\begin{figure}
    \centering
    \includegraphics[width=0.8\textwidth]{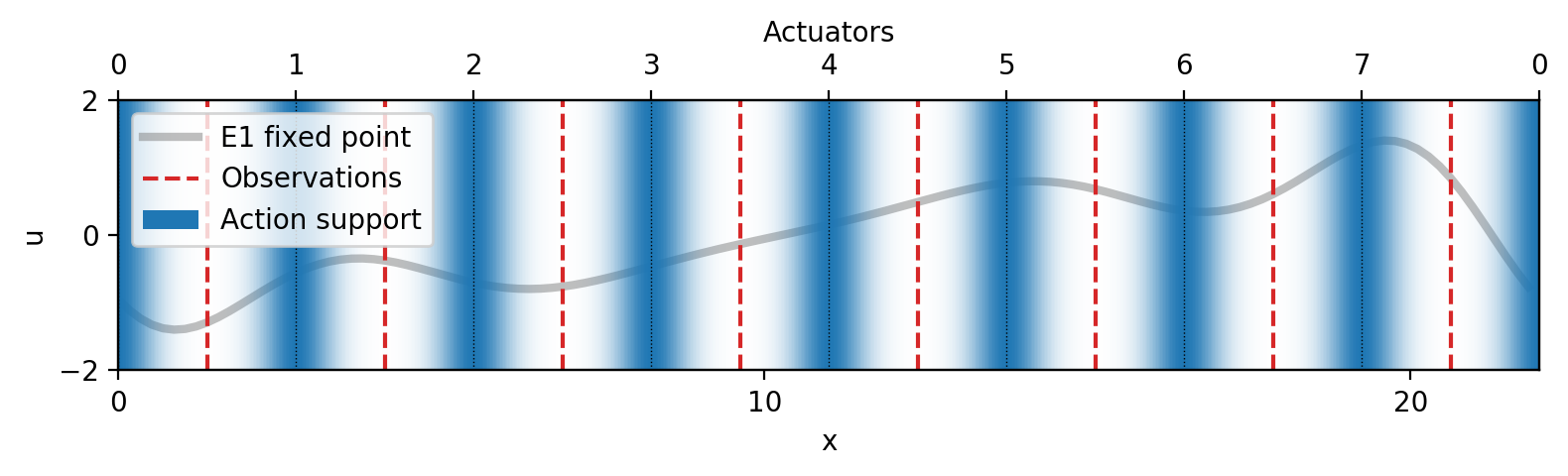}
    \caption{Location of observations (measuring $u$ at these locations) and Gaussian supports of the control actions overlaid on fixed point $E1$.}
    \label{fig:KS_layout}
\end{figure}

A control step is made of an update of the forcing action $a_t$, then $5$ time-steps and the measurement of the observations and reward. 
\red{This duration of $5$ numerical steps per control step is chosen as a trade-off with respect to two design criteria (as discussed by \citet{RabaultKuhnle2019}). The first is imposed by the flow dynamics, which involves characteristic evolution time-scales that need to be correctly observed by the agent. As measurements are performed at the end of each control step, $\Delta t$ must be short enough to discretise the observation signal properly and avoid aliasing for the agent to capture the flow dynamics. The second criterion is imposed by the training algorithm itself. The RL algorithm should be able to observe the impact of a given action in a short-term future (generally $<100$ control steps), which advocates for longer control-steps, that correspond to a further physical horizon.}

A standard run of the KS equation lasts for 250 control steps. The reset state is seeded using a Gaussian noise of amplitude $0.01$ and ran for a random number (from $40$ to $100$) of control steps without control action, so that control starts on a random state with a fully developed instability.

For the present study, the aim of the control is to stabilise $u$ around the fixed point $E_1$, and therefore the reward $r_t$ is defined as:
\begin{align*}
    MSE_t &= ||u(\cdot,t) - u_{E1}||_2 = \sqrt{\frac{1}{L}\int_0^L \left(u(x,t)-u_{E1}\right)^2 dx},\\
    r_t &= \frac{MSE_{ref}-MSE_t-0.1 ||\vec{a}_t||^2}{|MSE_{ref}|}
\end{align*}
where $u_{E1}$ describes the fixed point $E_1$ (refer to figure \ref{fig:KS_0} (right)), $MSE_{ref}$ is the time-averaged reference mean squared error of the uncontrolled state and $\vec{a}_t$ is the control action at time $t$. This way $r_t$ ranges over $]-\infty,1]$. \red{Note that the configuration being $x$-invariant, any translated version of $E_1$ is still a fixed point and could be considered a valid control target. Given the chosen reward, the system is here driven toward the specific fixed point $E_1$ from figure \ref{fig:KS_0} (right). Accounting for the $x$-invariance would require a more complex reward formulation, which is not explored in this paper.} 

\subsection{Results on the KS test-case}
\subsubsection{Pre-trained performances}
\begin{figure}
    \centering
    \includegraphics[width=0.7\textwidth]{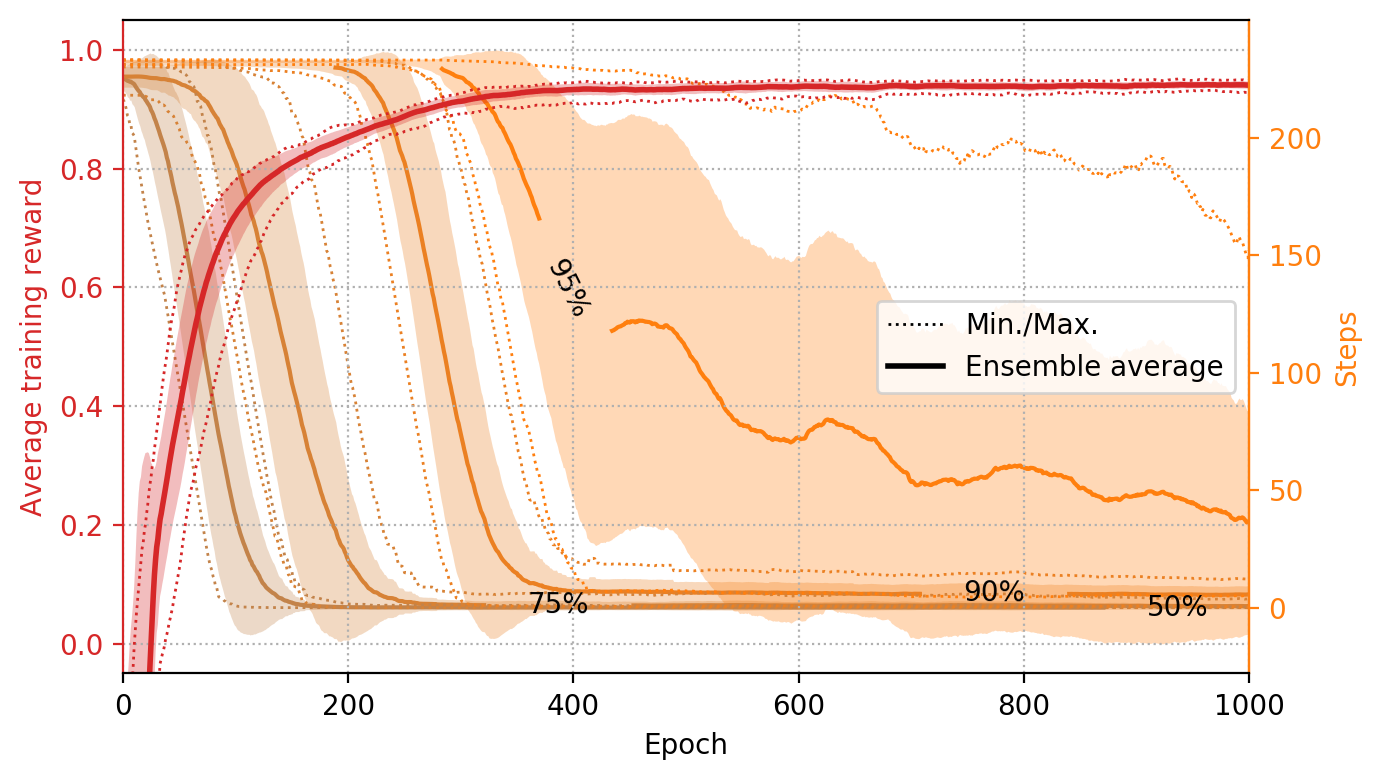}
    \caption{Average learning curve (red line) and evolution of the number of step needed to reach a given percentage (reported on the line) of the maximum performance (orange lines). Dotted lines describe the ensemble minimum and maximum and shaded areas illustrate the standard deviation across the batch. These averages are computed on a 10-case training batch. Orange lines have been smoothed using a rolling average of width 75 for the sake of readability.}
    \label{fig:KS_preTrained}
\end{figure}

\begin{figure}
    \centering
    \includegraphics[width=\textwidth]{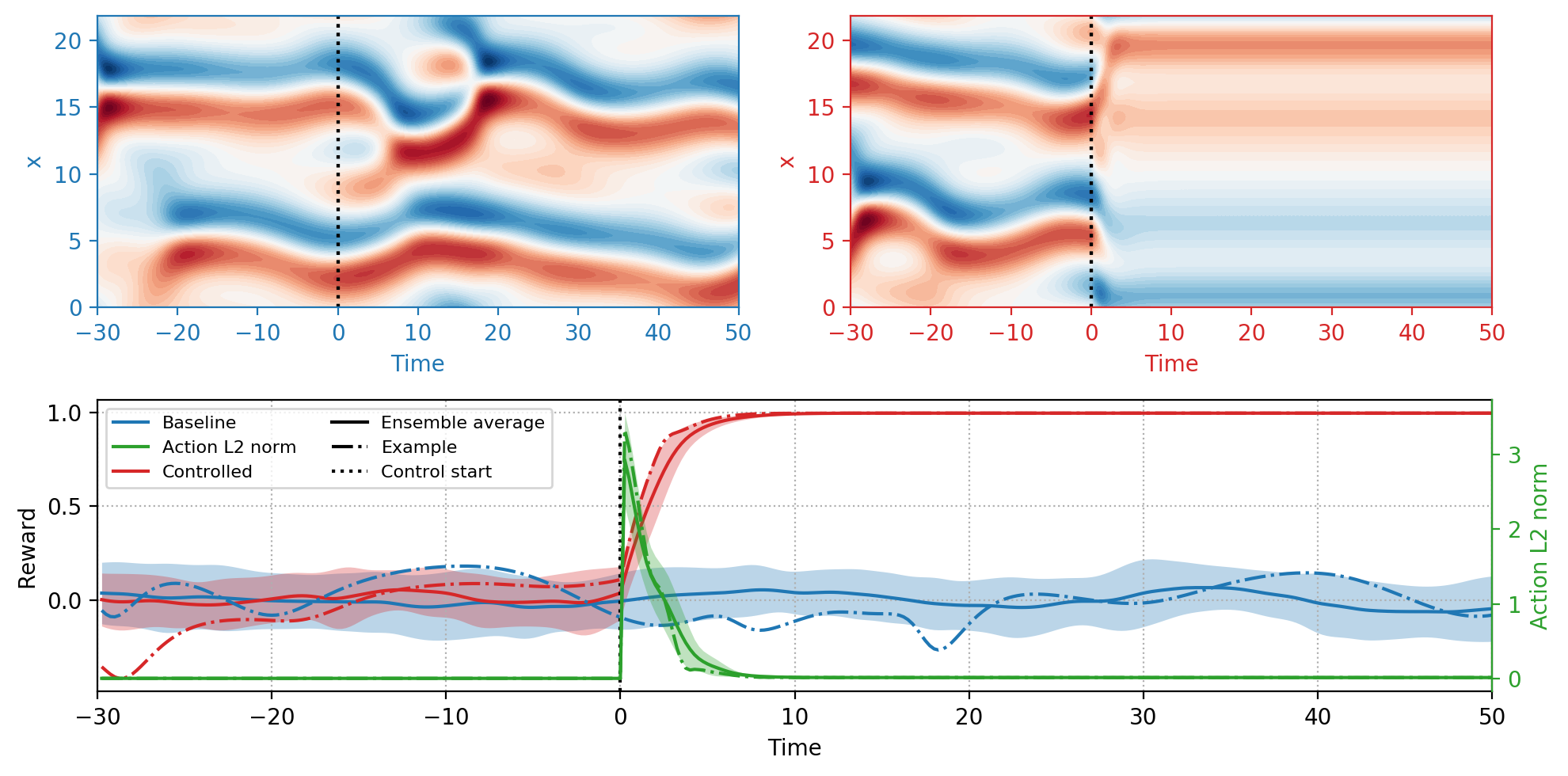}
    \caption{(bottom) Evaluation performances at the end of the pre-training. Averages are computed using 20 test runs. Normalised performance of the baseline (blue line) and of the controlled cases (red line) is to be read with the two left axes. Action $L_2$ norm (green line) is reported on the right axis. (top left and top right) Evolution of the state corresponding to the baseline and controlled ``Example" runs respectively.}
    \label{fig:KS_preTrained_ex}
\end{figure}

The initial training phase displays a steep increase of training performances until epoch $200$ as shown by figure~\ref{fig:KS_preTrained}. Normalised performance then slowly tends towards $0.95$. During this second phase the agent fine-tunes the policy in order to reach the target more precisely and to reduce the transient duration. Figure~\ref{fig:KS_preTrained_ex} shows ensemble averages of the performance and action norm during test runs at the end of the pre-training phase. One can clearly notice that the short transient from the baseline state to the target state is the consequence of large amplitude actions. Their amplitude decreases rapidly once the target state is reached.

\subsubsection{A systematic ablation study}\label{sec:KS_syst}
A first systematic study has been performed to get a reference in order to assess the performances of the proposed method. This study consists in training from scratch a policy with every possible actuator layout. This represents 255 different cases, each of these layouts being assessed using 5 test cases trained over 4000 epochs. Such an exhaustive study is only possible here because of the low computational cost of running the environment. 

It has been found that the optimal layout evolution is complex and incompatible with a one-by-one elimination strategy: for a given number of actuators $n_{act}$, the best layout is not necessarily a subset (a "child") of the optimal layout for $n_{act}+1$ actuators. Additionally, for a given $n_{act}$, the notion of "best layout" is subject to caution since several sub-optimal layouts display performances almost equal to the optimal. Overall, layouts with actuators evenly spread out in the domain demonstrate the best performances, without significantly differing from each others. 

These aspects make the comparison of the exhaustive ablation study with our method complex. \red{Thus it has been chosen to compare ensemble averages over the test batches rather than discussing slightly sub-optimal choices.}

\subsubsection{Sparsification process}\label{sec:KS_sparse}
\begin{figure}
    \centering
    \includegraphics[width=0.49\textwidth]{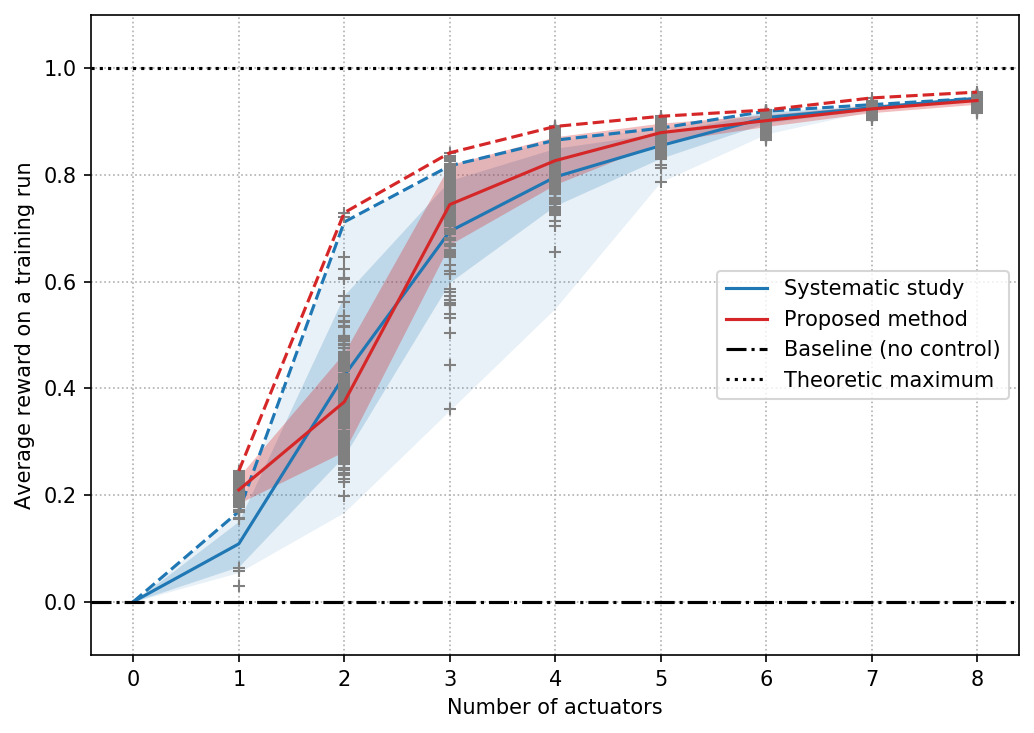}
    \includegraphics[width=0.49\textwidth]{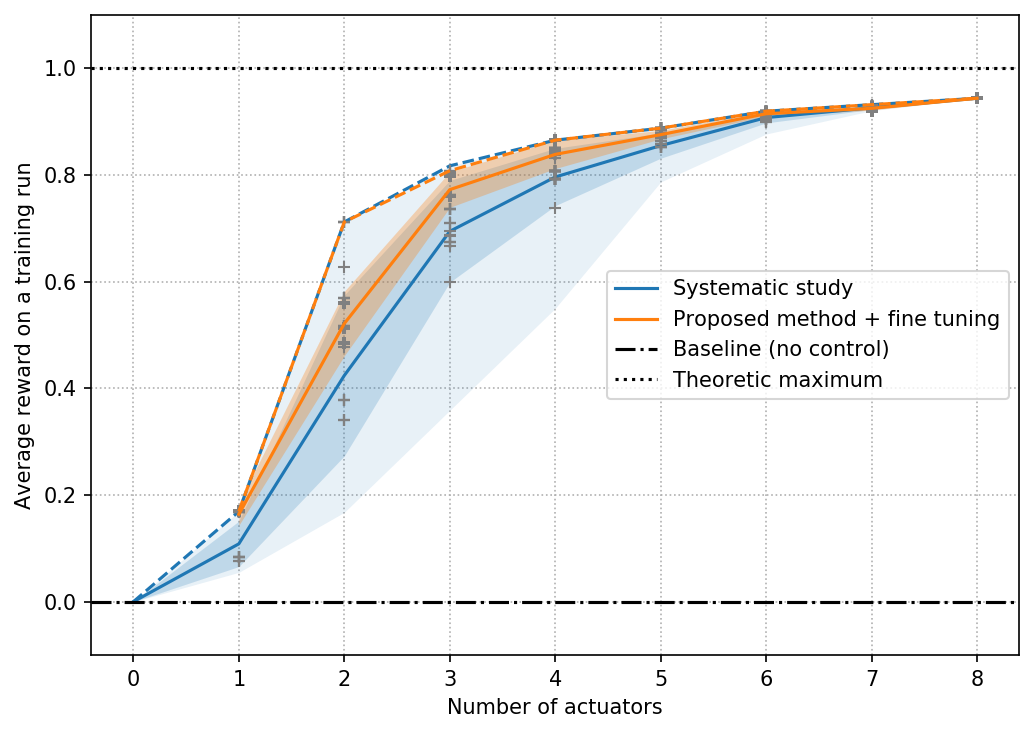}
    \caption{(left) Average training performance with respect to the number of active action components. The results of the proposed method (red line) are to be compared with the ones of the systematic study (blue line). The red shaded area illustrates the standard deviation across the batch, whereas the blue one represents the performance envelope of all possible layouts. Data points are denoted by grey "plus" signs. \red {(right) Expected performances of the proposed methods after an extra fine-tuning phase of the agents (orange line) compared to the systematic study (blue line).}}
    \label{fig:KS_sparse_perfo}
\end{figure}
The actuator elimination process is launched once the agent reaches a steady performance. The procedure described in section \ref{sec:2B_pract_impl} is implemented and the specific hyper-parameter values used for this case are reported in table \ref{tab:params} of the appendices. \red{Figure \ref{fig:KS_sparse_perfo} (left) synthesises the results of the sparsification process. The elimination of the first three actuators marginally impacts performance. However from 5 to 1 actuators, the average performance steadily decreases. A comparison with the results from the systematic study is also shown in the figure (blue lines). The obtained performance is below the average reward of the systematic study only for 2 actuators. This performance drop may be due to either a significantly sub-optimal layout selection from the algorithm (which would indicate a shortcoming of the propose algorithm) or unconverged policies on these newly selected layouts. To answer this point, figure \ref{fig:KS_sparse_perfo} (right) indicates the performances to expect if these obtained agents where to go through a further fine-tuning. This is done by considering the performance of the corresponding layout in the systematic study. One can notice that the average performance is higher with the proposed method than for the systematic study, even with only 2 actuators. The fact that fine-tuning would enable better-than-average performances with 2 actuators can be explained by the observed slow convergence of the performances of two-actuators layouts, which is considered as stable by the algorithm despite evolving slowing. A lower stability threshold may solve this issue but is likely to significantly slow down the elimination process.}

\red{Note that for 1 actuator, the average performance is slightly better than the optimal from the systematic study. One can assume that 1-actuator agents obtained by the elimination method take advantage of a broader exploration of the state-action space thanks to training phases with more than one actuator compared to the agents converged in the systematic study. This observation is further discussed in appendix \ref{app:KS_1act}.}


\begin{figure}
    \centering
    \includegraphics[width=\textwidth]{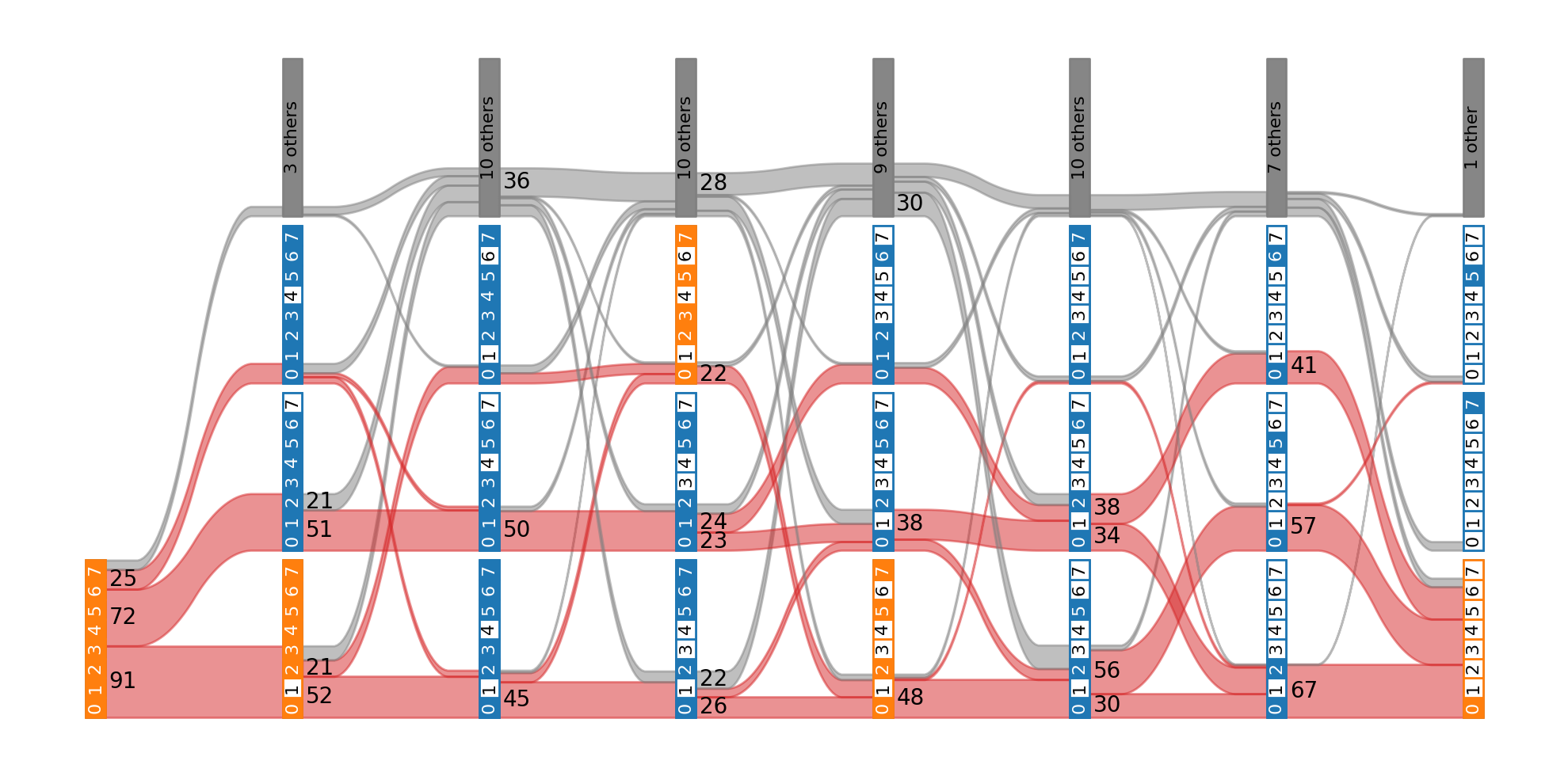}
    \caption{Detailed elimination history for the three main configurations for each complexity. \red{Stripes connecting layouts have a size proportional to the number of observed transitions (e.g. transitions from the 8-actuator layout to the 7-actuator layout having all active components but the last one, represent 72 out of the 200 test cases). Orange individuals denote overall best layouts for a given number of actuators.}}
    \label{fig:KS_cluster}
\end{figure}

Figure \ref{fig:KS_cluster} provides a more complete view of the elimination process by displaying the most frequent layouts obtained and the main paths towards sparsity. In a similar fashion as binomial combination coefficients, the number of layouts tends to increase toward 4 and 5 active components, but the three most selected layouts for each elimination step make up at least $65\%$ of the test batch. From 2 to 1 actuators, most of the individuals ($85\%$) converge toward the optimal 1-actuator layout. Orange-coloured layouts are the optimal ones for a given complexity, according to the systematic study. As pointed out before, these optimal layouts may not be ``children" of one another and may therefore be incompatible with our sequential elimination approach. This is one identified shortcoming of the present method: in the one-by-one elimination process, the algorithm makes a nearly optimal choice for a given number of actuators, but suffers from the constraint of its ``inherited" layout, restricting the choice of the next obtained layout.

\begin{figure}
    \centering
    \includegraphics[width=.49\textwidth]{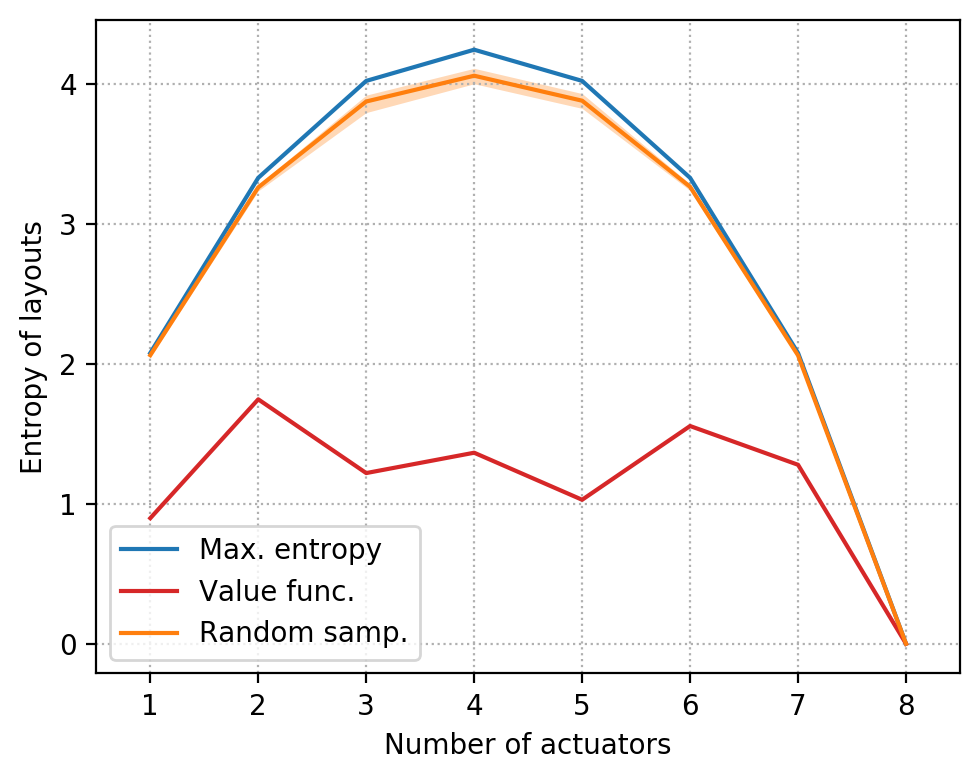}
    \includegraphics[width=.49\textwidth]{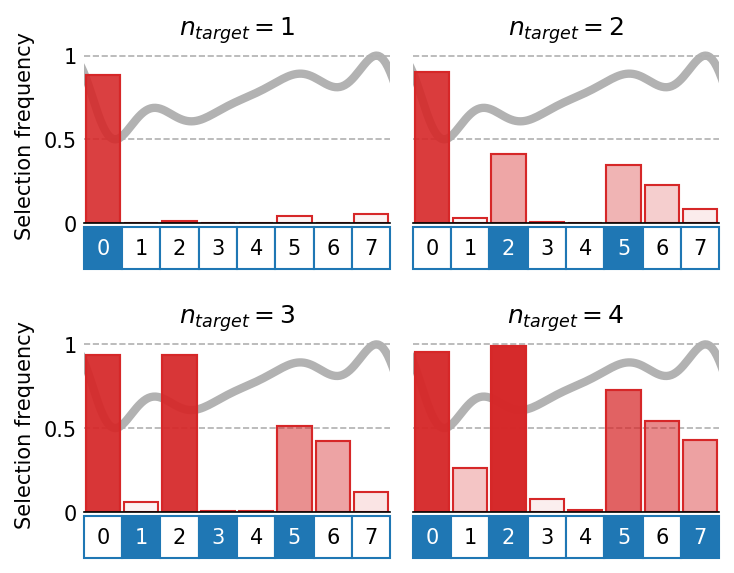}
    \caption{\red{(left) Entropy of the obtained layouts with respect to the number of active actuators. The maximum reachable entropy is the one reached by the systematic study (blue line). A random sampling of layouts of the same size as the study (200 test cases) would have the entropy represented by the orange line.} (right) Frequency histograms of the obtained layouts (bar plots) compared to the absolute best layout from the systematic study (blue numbered boxes) for 1 to 4 action components. Population sizes (on which frequencies are computed) are reported below each histogram. Fixed point $E_1$ has been plotted in the background as a reference.}
    \label{fig:KS_histo}
\end{figure}

Statistics about the obtained layouts are summarised in figure \ref{fig:KS_histo}. \red{The left-hand side graph shows the entropy $H$ of the layouts with respect to the number of actuators. $H$ is computed as follows:}
\begin{equation*}
    H = \sum_{l\,\in\,\text{layouts}} f_l \log\left(f_l\right),
\end{equation*}
\red{where $f_l$ is the observed frequency of a given actuator layout over the batch. The entropy obtained with the proposed algorithm (red line) compared to what an equivalent random sampling of the same size gives (orange line) demonstrates the selectivity of the method.} The right-hand side of the figure depicts the selection frequency of each action component for layouts having 1 to 4 active actuators. Here and in the following, actuators are numbered the same way as in figure~\ref{fig:KS_layout}. The elimination results show that actuators 3 and 4 are almost systematically eliminated for layouts of 4 or less action components, while actuators 0, 2 and, to a lesser extent, actuators 5 and 6 seem more important. 

Yet, the importance of a given action component is a relative notion since it depends on the context of the rest of the gated layout. For instance, actuator 1 may only be important for the performance if both actuators 0 and 2 are disabled, and less useful otherwise. Thus, the frequency analysis of figure \ref{fig:KS_histo} provides only a synthetic glimpse of the results, \red{but confirms that despite generating multiple layouts at some stages of the process, the method remains notably selective.}

\subsubsection{Partial conclusion on the study of the KS equation}
Despite being a computationally cheap, well-documented, one-dimensional test-case, the KS equation shows that even simple environments may present significant challenges for an actuator-selection algorithm. First, the KS equation seems to be controllable with a very small number of actuators (2) but only as long as they are properly spread out along the x-axis. This constraint is only partially overcome by the elimination method, mostly due to the difficulty to isolate the impact of one actuator from the rest of the layout (having active neighbours tends to reduce the estimated actuator's importance). Thus, starting from 8 actuators, the method has to perform 6 optimal (or nearly optimal since the order in which action components are removed \red{can be interchanged}) choices to reach the ``right" 2-actuator layout, leaving very little room for a sub-optimal -- yet efficient -- layout choice along the process. Second, this case may also highlight that simply switching a given action component off to assess its importance has limitations since it does not take the adaptation of the policy $\pi$ into account. 
This being said, starting from a given layout, the algorithm seems to perform the sequential elimination satisfactorily: despite relying on stochastic processes, the study shows that the algorithm has a high selectivity, and among the choice left by the sequential elimination paradigm, it yields layouts that will generally be among the high-performance ones.

\begin{table}
    \centering
    \begin{tabular}{ccc}
        Transition & $|\Delta V|/S(V_{syst.})$ (\%) & $S(V_{syst.})$\\
        \hline
        8 $\rightarrow$ 7 & $15.7\pm16.3$ & $0.003$ \\
        7 $\rightarrow$ 6 & $6.4\pm10.0$ & $0.012$ \\
        6 $\rightarrow$ 5 & $5.7\pm8.3$ & $0.024$ \\
        5 $\rightarrow$ 4 & $6.8\pm7.5$ & $0.054$ \\
        4 $\rightarrow$ 3 & $19.3\pm15.5$ & $0.096$ \\
        3 $\rightarrow$ 2 & $11.8\pm9.6$ & $0.151$ \\
        2 $\rightarrow$ 1 & $33.9\pm35.6$ & $0.043$ \\

    \end{tabular}
    \caption{Averaged relative value error $\Delta V = V_{estim.} - V_{obs.}$ for each actuator elimination. $V_{estim.}$ is the forecast value for the actuator elimination and $V_{obs.}$ is the observed value function after the elimination. $S(V_{syst.})$ is the standard deviation of the average performances of the all possible layouts (without taking any performance threshold into account).}
    \label{tab:KS_delta_perfo}
\end{table}

 To better characterise the performance reached by the algorithm, one may compare the difference of estimated state values that drive the algorithm with the real values computed from the actual returns \red{after elimination}. Table \ref{tab:KS_delta_perfo} uses $S(V_{syst.})$ (the standard deviation of the performance of all possible layouts from the systematic study) as a reference value to quantify the required level of accuracy needed to properly estimate the value of a given actuator elimination. Should the estimation error $\Delta V=V_{estim.} - V_{obs.}$ be close to $S(V_{syst.})$, the estimation would not be precise enough to reliably select the "right" actuator.
 \red{The highest error is encountered for transition 2 $\rightarrow$ 1 and makes up around $1/3$ of the typical dispersion of values. Despite this large estimation error, most of the elimination choices lead to the optimal 1-actuator layout as explained before. Then come transitions 4 $\rightarrow$ 3 and 8 $\rightarrow$ 7. The first may come from the adaption of the policy, which is not taken into account by the estimated value $V_{estim.}$. The latter is likely due to the very low dispersion of $V_{syst.}$ values (eliminating any actuator has nearly the same impact when all the others are kept active). For all the transitions, $V_{estim.}$ under-estimates $V_{obs.}$ ($\Delta V<0$ on average) as one can reasonably expect. But overall, one may see that the metric used for the actuator elimination is rather well estimated by the algorithm, showing that the resulting sub-optimal choices appear to come from the complexity of the KS case regarding optimal actuator placement. This case allows to pinpoint some interesting improvement directions for future work, discussed in the conclusion.}
 

\renewcommand*{\MyPath}{P3_NACA}
\renewcommand*{\img}{P3_NACA/img}
\section{Low-Reynolds NACA 0012 stalled flow}\label{sec:NACA}

\begin{figure}
    \centering
    \raisebox{-0.5\height}{\includegraphics[page=23,width=0.4\textwidth]{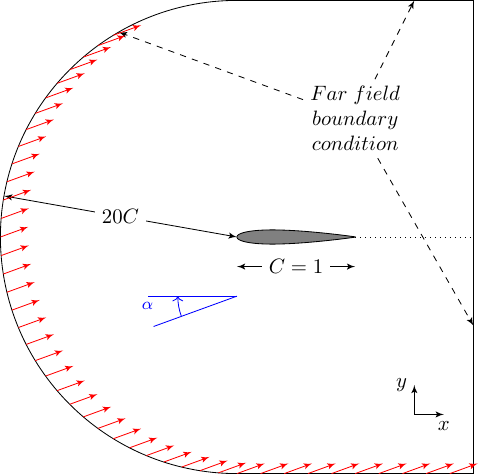}}
    \hspace{2em}
    \raisebox{-0.5\height}{\includegraphics[page=36,width=0.35\textwidth]{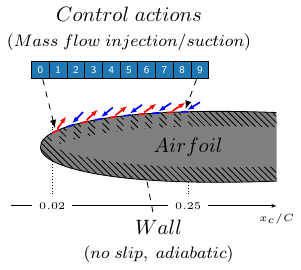}}
    \caption{(left) Flow domain geometry, not at true scale. $\alpha$ denotes the angle of attack and $C$ is the (unitary) chord length. (right) Boundary conditions on the airfoil, with $n_{act}=10$. Blue numbered boxes symbolise actuators, number 0 being the upstream-most actuator, number 9 being the downstream-most one.}
    \label{fig:NACA_domain}
\end{figure}

The second test case concerns the control of a bi-dimensional flow around a stalled NACA 0012 at a chord-wise Reynolds number $Re_c=1000$. 
As illustrated by figure \ref{fig:NACA_domain}, the origin of the domain $(0,0)$ is located at the airfoil leading edge, the computational domain is "C-shaped" and extends up to a distance of $20$ chord lengths ($C=1$). The simulation is built in the reference frame of the airfoil, meaning that the angle of attack ($\alpha$) is imposed by the upstream flow conditions, modelled here as a far-field boundary condition.

The free-stream flow is uniform at $M_\infty=0.1$. In the following, all quantities are made non-dimensional by the characteristic length $C$, the inflow density $\rho_\infty$, the velocity $U_\infty$ and the static temperature $T_\infty$. The flow solution is computed via direct numerical solving \red{(no turbulence modelling)} using ONERA's FastS finite volume method solver \citep{Dandois2018}. The spatial discretisation is based on \red{a simplified version \citep{Mary1999} of a} second-order-accurate AUSM+(P) scheme proposed by \citet{Edwards1998}, and the time-integration of the equations is performed with a second-order implicit Euler scheme, with a global non-dimensional time step $dt~=~1.3\times~10^{-3}$. The structured mesh is made of $120,000$ nodes distributed such that the vicinity of the airfoil and in its wake are properly resolved. The boundary conditions are specified in figure \ref{fig:NACA_domain}.

A control step $\Delta t$ lasts for $58$ numerical time steps of the simulation, corresponding to $\Delta t= 7.9\times10^{-2}$ time units. \red{Similarly to the KS test-case, this choice was made to both comply with the needs for a long enough horizon for the agent and for a precise enough discretisation of the measured signal.} 
Here $\Delta t$ corresponds to $\approx1/50$ characteristic period, assuming that most of the consequences of a control action are measured within the next two characteristic periods.

\begin{figure}
    \centering
    \includegraphics[width=0.8\textwidth]{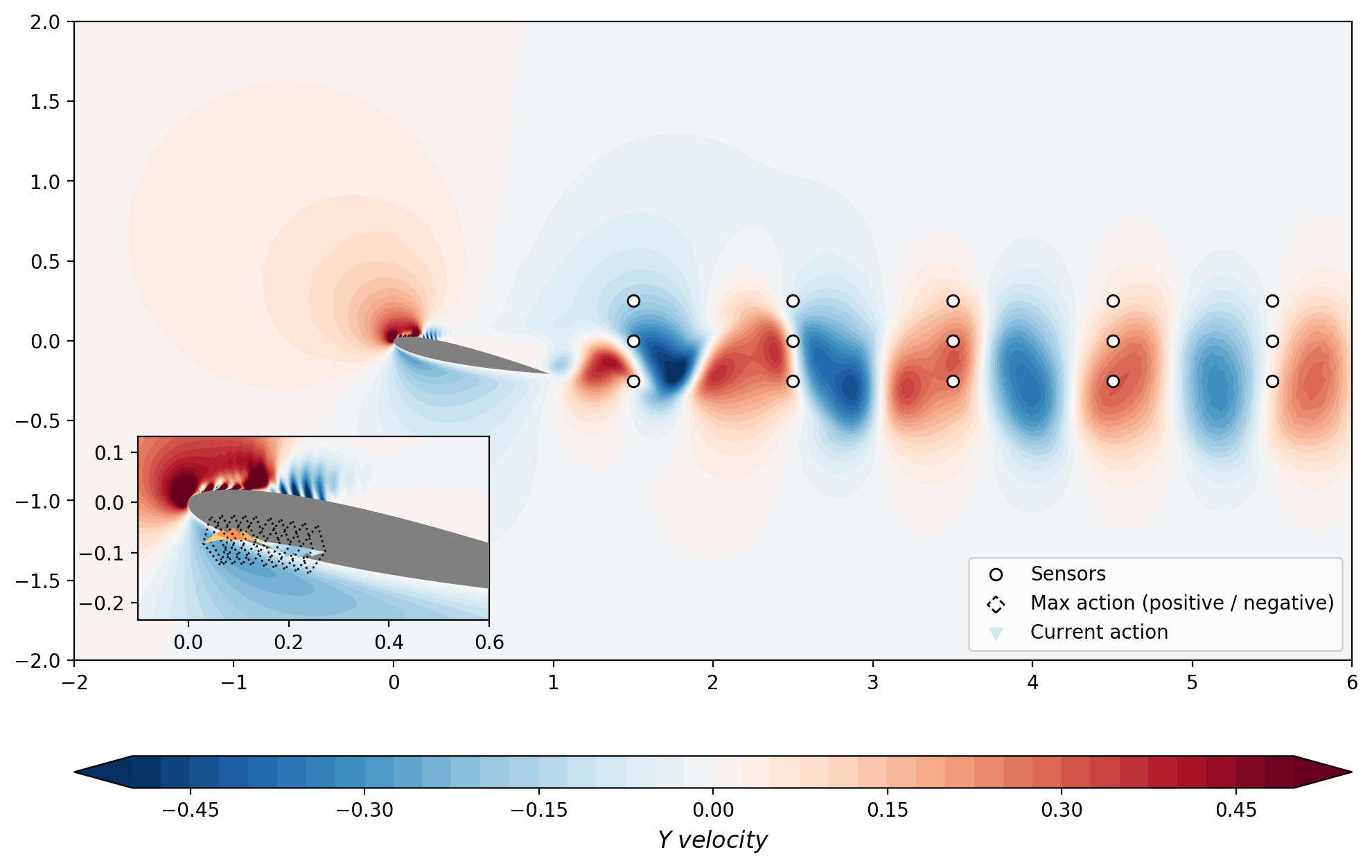}
    \caption{Instantaneous $Y$ velocity flow field, with an arbitrary control action (here with $n_{act}=20$), in the \textbf{free-stream reference frame}. White dots represent the sensor locations. The coloured triangles nearby the airfoil depict the action, their heights and colours representing each action amplitude. The dashed diamond shapes mark off maximum actions (both positive and negative). The strong variations in velocity in the vicinity of the actuators are due to the presence of interspersed wall boundary conditions in-between actuators.}
    \label{fig:NACArefCase}
\end{figure}

Control action is performed on the airfoil suction side through a series of $n_{act}$ independent jet inlets. Negative control actions correspond to suction and positive to blowing, the later being performed at an angle of $-80$° with respect to the local wall normal. The control action (ranging $[-1,1]^{n_{act}}$)  is first re-scaled before each control step to the actuators' action limits (here $\pm2$) and converted to a mass-flow setpoint for the current control step. This command is used to compute the mass-flow boundary condition imposed for each time-step using a $52$-iteration interpolation ramp \red{($90\%$ of the step)} in order to avoid abrupt changes that may not be handled by the numerical solver, in a similar fashion as \citet{ParisBeneddineDandois2021} and \citet{RabaultKuchtaJensenEtAl2019} did. Thus, for the $i^{th}$ numerical iteration of control step $t$, the mass-flow per unit area $q_j^i$ implemented for actuator $j$ reads:
\begin{align*}
    q_j^i = \rho_\infty U_\infty\left(a_{j,t-1}^i(1-r_i) + a_{j,t}^i r_i\right), \text{ with }   r_i = \min(i/52,1).
\end{align*}
The stagnation enthalpy is held constant (at the upstream flow value) for injection actions. Figure \ref{fig:NACArefCase} illustrates a standard setup for this case.

Both drag and lift coefficients ($C_d$ and $C_l$) are computed on the airfoil via the resulting force of the flow $\vec{F}$:
\begin{equation}
    \vec{F} = \begin{pmatrix}F_x\\F_y\end{pmatrix} = \oint_{\text{airfoil}}\mat{\sigma}.{\vec{n}}dS,
\end{equation}
\begin{equation}
    C_d = \frac{1}{\frac{1}{2}\rho_\infty U_\infty^2 C}\begin{pmatrix}\cos\alpha&\sin\alpha\end{pmatrix}\begin{pmatrix}F_x\\F_y\end{pmatrix},
\end{equation}
\begin{equation}
    C_l = \frac{1}{\frac{1}{2}\rho_\infty U_\infty^2 C}\begin{pmatrix}-\sin\alpha&\cos\alpha\end{pmatrix}\begin{pmatrix}F_x\\F_y\end{pmatrix},
\end{equation}
where $\vec{n}$ is the unitary airfoil surface normal vector, $\mat{\sigma}$ is the stress tensor, $\vec{e}_x = (1,0)$, $\vec{e}_y = (0,1)$ and $\alpha$ is the angle of attack of the airfoil. Note that lift and drag coefficients are computed by integration around the airfoil on a closed circulation, in the presence of actuators.

\subsection{Non-controlled flow}
\begin{figure}
    \centering
    \includegraphics[width=\textwidth]{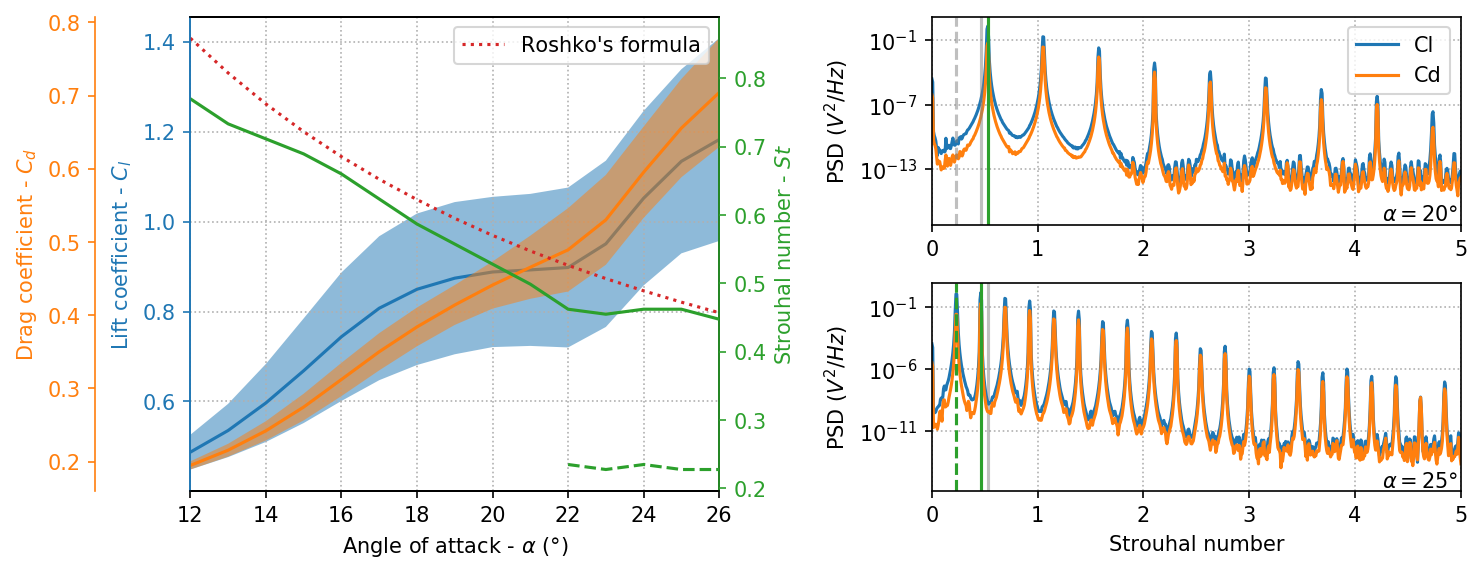}
    \caption{(left) Evolution of the lift and drag coefficients and of the Strouhal number with the angle of attack $\alpha$. The shaded areas depict the standard deviation of $C_l$ and $C_d$. The secondary Strouhal number emerging for $\alpha\geq22$° is drawn in a dashed line. The red dotted line shows the predicted Strouhal number for an equivalent cylinder using Roshko's formula. (right) $C_l$ and $C_d$ power spectral density (PSD) spectra plotted for $\alpha=20$° (top) and $\alpha=25$° (bottom). Green vertical lines denote the main Strouhal numbers. Other Strouhal numbers (for a the other value of $\alpha$) are reported as grey lines for comparison.}
    \label{fig:NACA_no_ctrl_0}
\end{figure}

In this study, the angle of attack $\alpha$ ranges from $12$ to $20$ degrees. At $Re_C=1000$, the flow is unsteady and displays a laminar vortex shedding \citep{WuLuDennyEtAl1998}. This instability causes both lift and drag coefficients to vary periodically, yielding undesired alternated loads on the airfoil. The evolution of both the amplitude and the periodicity of these loads are shown in figure \ref{fig:NACA_no_ctrl_0}. The left-hand side graph shows evolution of the lift and drag coefficients with respect to the angle of attack $\alpha$. The Strouhal number decreases up to $\alpha=22$° and remains nearly steady for larger $\alpha$ values, in agreement with Roshko's correlation formula for a low-Reynolds cylinder \citep{Roshko1953}. The latter is computed considering the cross-sectional "area" of the airfoil facing the air flow as diameter of an equivalent cylinder. The formula is used in its range of validity since the diameter-based Reynolds number $Re_{\emptyset}$ ranges from $233$ ($\alpha=12$°) to $451$ ($\alpha=26$°).

Figure \ref{fig:NACA_no_ctrl_0}~(right) depicts both lift and drag coefficient spectra for angles of attack of $20$° and $25$°. For $\alpha=20$°, both lift and drag coefficients have a very peaked harmonic spectrum with a fundamental Strouhal number $St(\alpha=20)\approx0.52$. For $\alpha=25$°, interspersed peaks appear in-between pre-existing harmonics. The main Strouhal number slightly decreases to $St(\alpha=25)\approx0.46$ and a halved Strouhal number $St'$ can be measured at around $0.23$. The emergence of the latter for $\alpha\geq22$°, whose evolution is reported on the left-hand side graph (green dashed line) causes to break the symmetry between two successive shedded vortex pairs.

\begin{figure}
    \centering
    \includegraphics[width=0.8\textwidth]{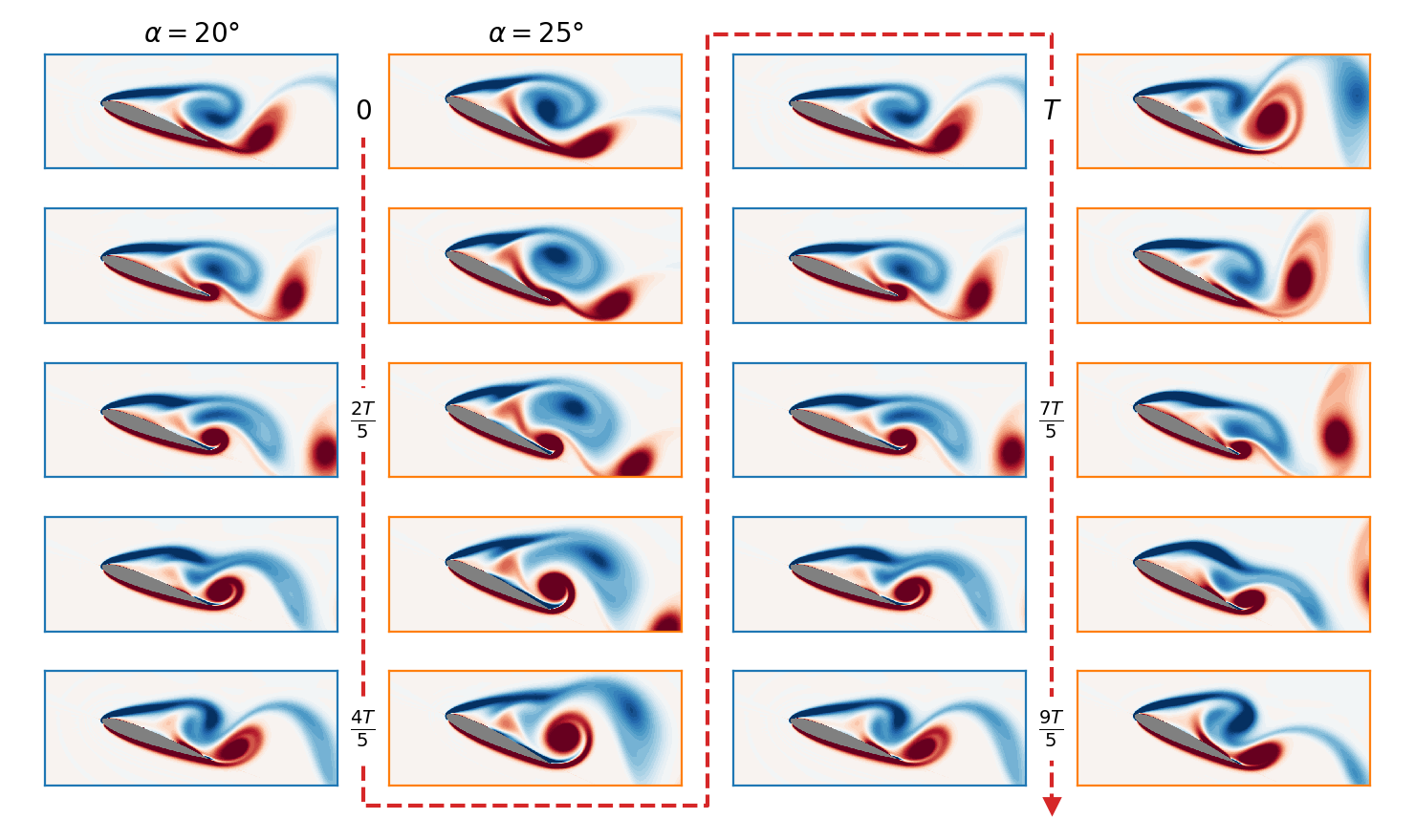}
    \caption{Snapshots of the flow vorticity around the airfoil over two characteristic periods $T$. Blue-framed thumbnails correspond to $\alpha=20$°, orange-framed ones correspond to $\alpha=25$°.}
    \label{fig:NACA_no_ctrl_1}
\end{figure}

Figure \ref{fig:NACA_no_ctrl_1} illustrates this symmetry breaking over two vortex periods. During the first period, the negative-vorticity, suction-side vortex (coloured in blue) induces the shedding of a weak pressure-side vortex downward and feeds a second positive-vorticity vortex on the suction side, near the trailing edge. At the end of the first period, the latter induces an upward shedding of this clockwise-rotating vortex. During the second period, the trailing-edge vortex is stronger and higher compared to its counterpart one period before. It sheds horizontally and stretches the leading-edge vortex induced. Compared to the case $\alpha=20$°, where trailing-edge and leading-edge vortices have comparable sizes and strengths, at $\alpha=25$°, the leading-edge then trailing-edge vortices are alternatively dominant over time. 

For any angle of attack $\alpha$, one can define the characteristic period $T=1/St(\alpha)$ of the unstable phenomenon. This time unit is later used in the study to size the control step. The main goal of the controller is to minimise lift fluctuations using as little control power as possible. Thus, the reward $r_t$ is defined as:
\begin{align*}
    r_t = - S\left(C_l\right)_{2T} - S\left(C_d\right)_{2T} - 0.05\frac{1}{n_{act}}\sum_{i=0}^{n_{act}-1}|\langle a_i\rangle_{2T}|,
\end{align*}
where $S\left(C_l\right)_{2T}$ and $S\left(C_d\right)_{2T}$ are the standard deviation of the lift and drag coefficients computed over two characteristic periods and $|\langle a_i\rangle_{2T}|$ the absolute value of the averaged $i^{th}$ action component also over 2 periods.

\begin{figure}
    \centering
    \includegraphics[width=0.5\textwidth]{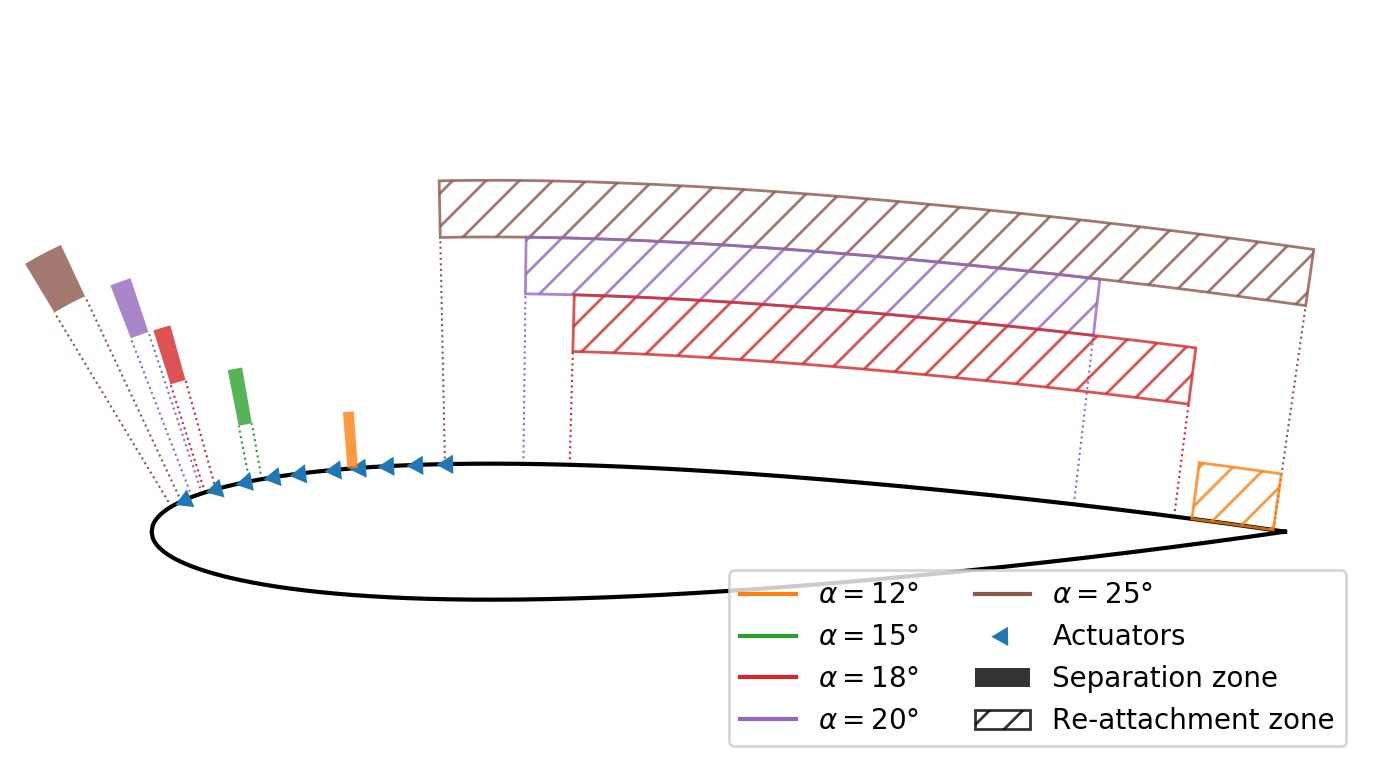}
    \caption{Separation (filled areas) and re-attachment (hatched areas) points ranges with respect to the angle of attack. For $\alpha=15$°, the re-attachment takes place at the trailing edge.}
    \label{fig:NACA_sep}
\end{figure}

Figure \ref{fig:NACA_sep} describes the ranges of both separation and re-attachment points with respect to the angle of attack. The separation point ranges between $17.3$\% and $18.2$\% (of the chord length) for $\alpha=12$°, between $8.6$\% and $9.7$\% for $\alpha=15$° and between $3.6$\% and $4.4$\% for $\alpha=20$°. This expected reduction of the chord length of the separation comes along with a strong variation of the re-attachment point due to a growing vortex-shedding unsteadiness with $\alpha$ as shown by the figure. For $\alpha=15$°, no re-attachment is measured, but one can consider it is located at the trailing edge.
\subsection{Results on the NACA test-case}
The proposed algorithm has been run on test-cases with the previously introduced environment setup for angles of attack ($\alpha$) of $12$°, $15$° and $20$°. The choice of these three cases is motivated by the significantly different dynamical behaviour they exhibit. The specific hyper-parameter values used for this study are reported in table \ref{tab:params}. For the sake of conciseness, comparison of the results between these cases is only presented and discussed when noticeable differences arise.

\subsubsection{Pre-trained performances}
\label{sec:NACApretrained}
\begin{figure}
    \centering
    \includegraphics[width=\textwidth]{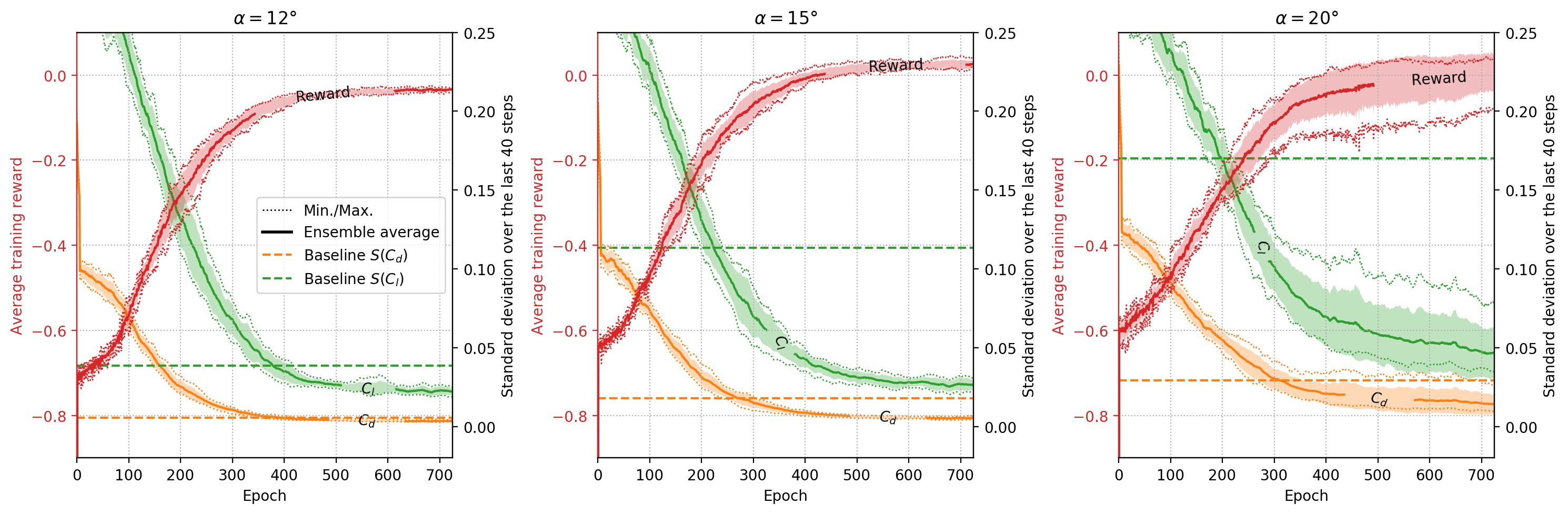}
    \caption{Ensemble-averaged learning curves of the pre-training phase for $\alpha=12$° (left), $\alpha=15$° (centre) and $\alpha=20$° (right). Dotted lines describe the ensemble minimum and maximum and shaded areas illustrate the standard deviation across the batch.}
    \label{fig:NACA_preTrained_lc}
\end{figure}

For each angle of attack, ten independently initialised test cases have been trained. Figure \ref{fig:NACA_preTrained_lc} illustrates the evolution of both the average training reward and the standard deviation of the lift ($C_l$) and drag ($C_d$) coefficients at the end of each training run. During the pre-training phase (from epoch 0 to 700) and for all of the cases, the average reward grows rapidly until epochs $300\sim400$ then stabilises to its maximum value. This comes with a steep reduction of the variations $C_l$ and $C_d$. The average exploration component $\sigma$ of the policy steadily decreases during training (not shown). This behaviour is expected from the agent, that starts with a broad exploration of the possible actions and then narrows it down towards a more deterministic control strategy.

\begin{figure}
    \centering
    \includegraphics[width=\textwidth]{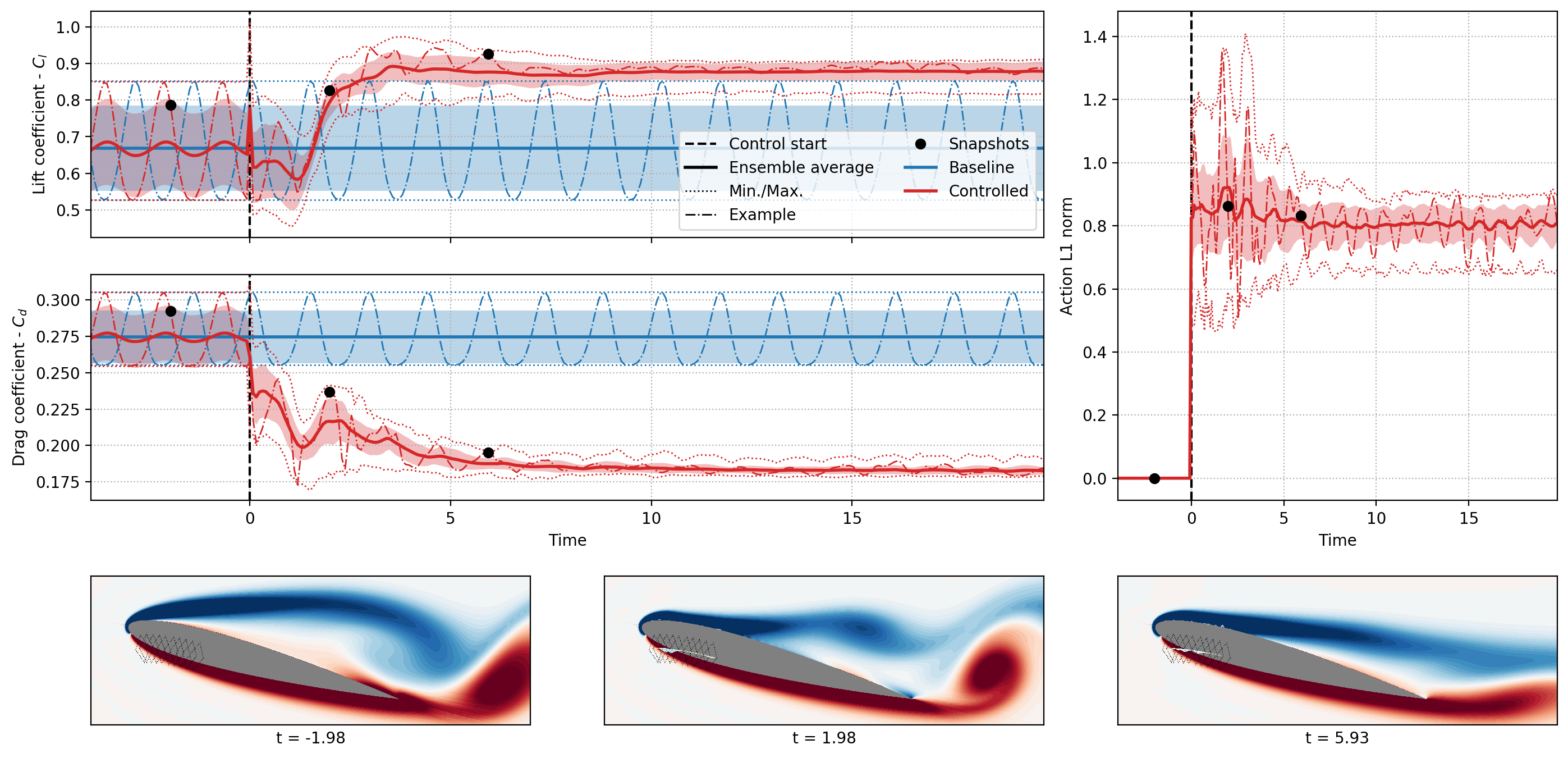}
    \caption{Ensemble averages of evaluation runs performed at the end of the pre-training phase (epoch $700$) for $\alpha=15$°. Here $10$ independent test-cases are evaluated on 10 test-runs (100 trajectories in total). Dotted lines describe the ensemble minimum and maximum and shaded areas illustrate the standard deviation across the batch. Vorticity snapshots illustrate the flow state at key moments of a randomly chosen run.}
    \label{fig:NACA_preTrained_eval15}
\end{figure}

Figure \ref{fig:NACA_preTrained_eval15} shows the ensemble averages of evaluations performed at the end of this pre-training phase for $\alpha=15$° and vorticity snapshots of the flow along a randomly selected test-run. As soon as the control starts, the control action swiftly drives the lift and drag coefficients towards favourable and stable values. As shown by both the extremal envelopes and the flow snapshots, this correlates with a stationarisation of the flow through the cutoff of vortex shedding and with flow re-attachment. Control actions are moderate suction actions as expected. In the following, actuators are numbered according to figure \ref{fig:NACA_domain}.

Contrary to the KS case, a systematic study to compute optimal layouts for the airfoil cases is not possible due to the \red{unaffordable computational resources it would require}. But this flow configuration is nonetheless interesting because the control laws are simple enough such that a first analysis of each action component provides insights on the usefulness of each actuator. This makes a relevant source of information to evaluate the sparsification algorithm performance. The following paragraph provides the main conclusions obtained from a careful observation of each action components for the converged policies. The figures supporting the following claims are provided in appendix \ref{app:NACA}.

For $\alpha=20$°, all action components display strong variations during the transient phase which lasts for about $8$ non-dimensional time units. They all remain unsteady afterwards, synced on the vortex shedding that has not been entirely cancelled. Action components 0 and, to a lesser extent 1, 2 and 3, display a strong suction control forcing after the transient. These forcings ensure the flow re-attachment. 
For cases with $\alpha=15$°, actuators 3, 4 and 6 enforce fast varying actions. The re-attachment is again ensured by actuators 0 and 3, and to a lesser extent, actuator 1. It is worth noting that after the transient (lasting for around $6$ non-dimensional time units), all the action components excepted actuator 6 become relatively stable.
At last, for $\alpha=12$°, only actuator 3 has a strong action variation during the transient of duration of around $5$ non-dimensional time units. During the stabilised phase, action components 2, 3 and 7 display a strong and relatively steady suction forcing. Again here actuators 6 and 9 remain relatively unsteady compared to the others.

\subsubsection{Sparsification process}
\begin{figure}
    \centering
    \includegraphics[width=\textwidth]{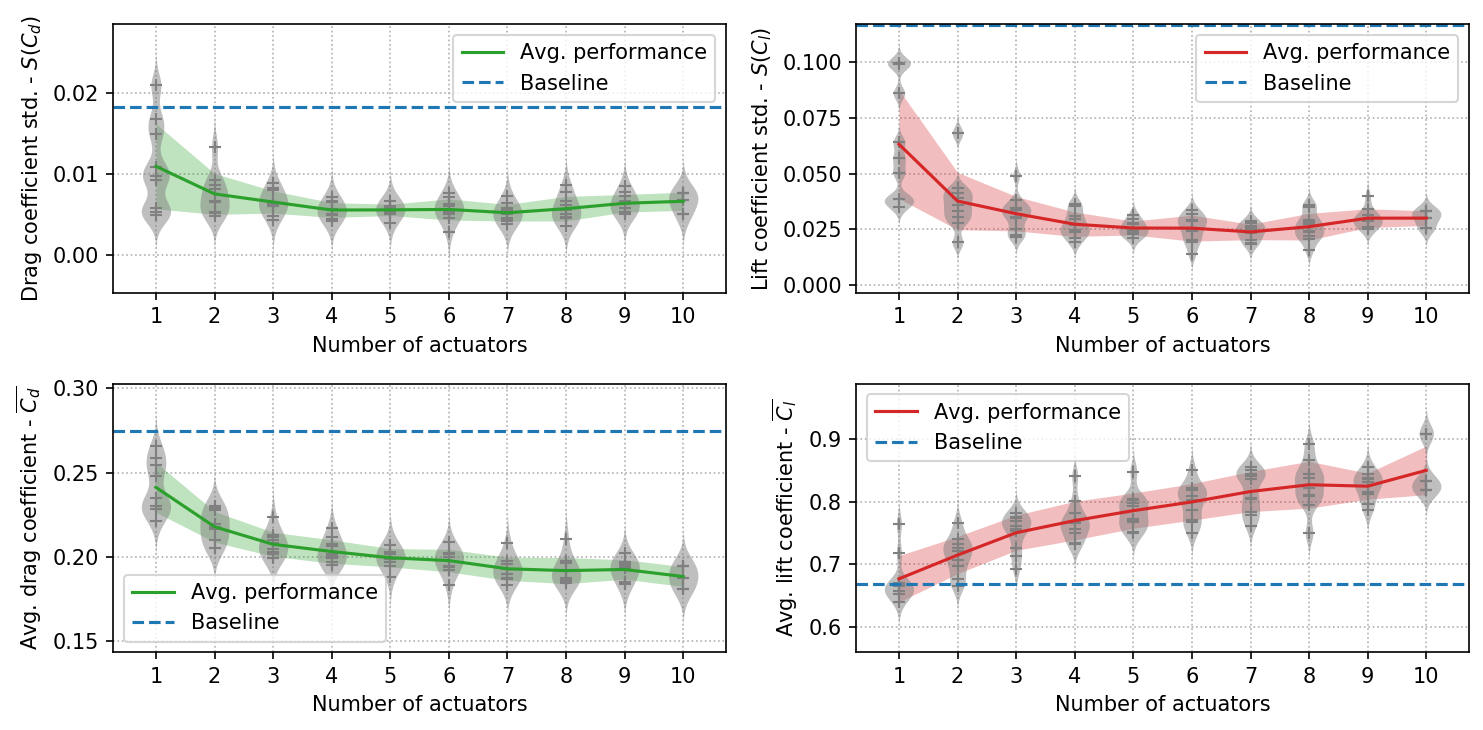}
    \caption{Averaged training performance indicators with respect to the number of active action component for $\alpha=15$°. Data points are denoted by grey ``plus" signs. Ensemble averages are computed on these data points (batch-size $5$). The blue dashed line represents the uncontrolled (baseline) performance, solid lines denote the evolution of the ensemble average and shaded areas illustrate the standard deviation across the batch. All indicators are computed on the last 40 control steps of a training run. (Upper left) Evolution of the standard deviation of the drag coefficient $S(C_d)$. (Upper right) Evolution of the standard deviation of the lift coefficient $S(C_l)$. (Lower left) Time-averaged drag coefficient $\overline{C_d}$. (Lower right) Time-averaged lift coefficient $\overline{C_l}$.}
    \label{fig:NACA_sparse_perfo_15}
\end{figure}

Figure \ref{fig:NACA_sparse_perfo_15} synthesises the results of the sparsification on the cases with $\alpha=15$°. Since performances follow similar trends for $\alpha=12$° and $\alpha=20$°, the sparsification process is discussed for $\alpha=15$° only. One can distinguish two different trends in the performances. First, from 10 to 4 actuators, only the average values of both drag ($\overline{C_d}$) and lift ($\overline{C_l}$) coefficients (refer to lower left and right graphs of figure \ref{fig:NACA_sparse_perfo_15}) are significantly impacted by the elimination of actuators, respectively seeing an increase and a drop of their value as the number of actuators is reduced. Standard deviations of these coefficients (on which the reward is based), quantifying the steadiness of the mechanical loads, remain rather constant. As the number of actuators further decreases, from 4 to 1, both average coefficient values pursue and amplify their previously described evolution. Standard deviations significantly increase, denoting an expected decrease in overall performance. For the vast majority of the data points, the control yields significantly better performances compared to the non-controlled flow (denoted as ``baseline" on figure \ref{fig:NACA_sparse_perfo_15}). 

\begin{figure}
    \centering
    \includegraphics[width=\textwidth]{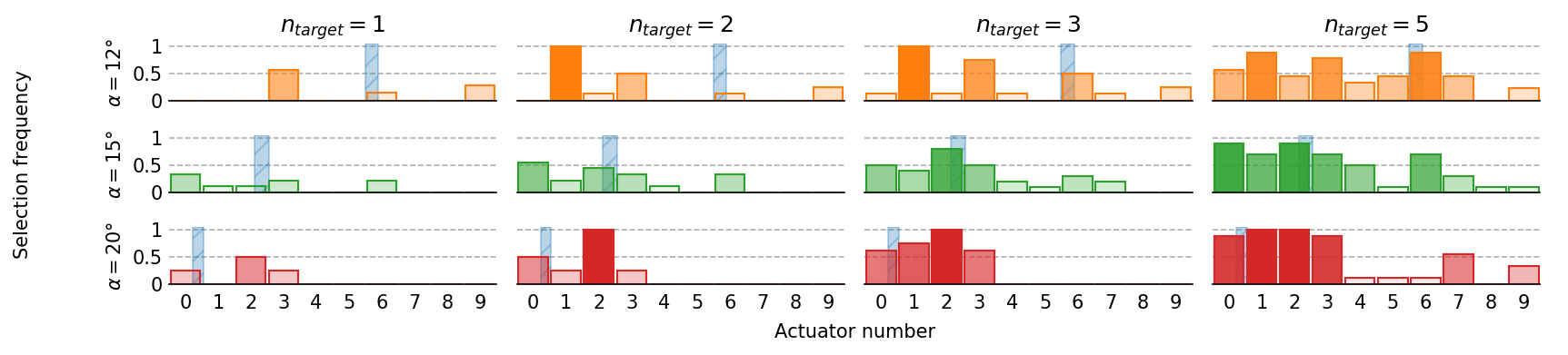}
    \caption{Frequency histograms of the obtained layouts (bar plots), for $\alpha=12$° (top row), $\alpha=15$° (middle row) and $\alpha=20$° (bottom row). The blue shaded areas correspond the range of evolution of the separation point. These frequencies are observed on the test batch of size 10, used in this study.}
    \label{fig:NACA_sparse_hist_comp}
\end{figure}

Figure \ref{fig:NACA_sparse_hist_comp} further describes the elimination choices with respect to both the angle of attack $\alpha$ and the number of active action components. In all cases, the algorithm chooses actuators consistently to the first analysis proposed in the end of section \ref{sec:NACApretrained}. With 5 and 3 components, eliminations for all studied angles of attacks display similar patterns, removing actuators 4 and 5, preserving other actuators with a slight advantage for upstream ones (0 to 3). For $\alpha=15$° and $\alpha=20$° and for a reduced number of action components (1, 2 and 3), the algorithm favours locations near the separation range, as one may intuitively expect. However, for $\alpha=12$° the remaining actuators' location differs from the separation range. This turns out to be more beneficial for this specific angle of attack, as shown in appendix \ref{app:NACA} (figure \ref{fig:NACA_12deg_upst_vs_dnst}).

\subsubsection{Partial conclusion on the study of the NACA flow}
The study of this test case has first shown that, except for $\alpha=20$°, a nearly complete stabilisation of the flow using reduced-amplitude control actions, could be obtained with the proposed layout and using the described reinforcement learning method. The proposed action sparsification algorithm has shown expected results while still allowing to significantly preserve control performances during the elimination phases. For reduced numbers of action components, the feedback provided by the sensors enables the policy to properly synchronise control actions with the remaining unsteadiness and thus, to reduce effective load variations on the airfoil.

\subsubsection{Remark on the obtained closed-loop control laws}
The airfoil case has been selected because of the rather simple control policies it yields (in most of the cases, a suction in the vicinity of the separation region), which makes it a good candidate to assess the behaviour of our actuator elimination algorithm.
As the control actions appear nearly steady during the late phase of control (whether it is with the full 10-actuator layout or the sparse ones), one may question the intrinsic closed-loop nature of the control law. Note that this question is not in the original scope of the present paper. It is nonetheless an interesting side point that is briefly addressed in the following. 

Closed-loop tests runs have first been performed for $10$, $5$ and $2$ actuators for all considered angles of attack, to be used as baseline values. Action sequences have then been re-implemented in an open-loop fashion on randomly reset cases. Table \ref{tab:NACA_ol_vs_cl} presents the variation of various performance indicators at the end of the test runs.

Relative variations of the time-averaged lift coefficient are systematically positive, whereas the drag coefficient evolves negatively. These variations yet remain small compared to the absolute coefficient values. Standard deviations quantifying the steadiness of the loads on the airfoil (which constitute the actual control objective defined in the reward) are systematically increased when implementing control action in an open-loop fashion. For 10 actuators, the relative variation showing an important increase must be relativised knowing that the reference standard deviation is very low, thus inflating the importance of the variation. The difference between open-loop and closed-loop can be seen for $5$ and $2$ actuators, where the closed-loop does not totally stabilise the flow. Thus, here it can be considered that for 10 actuators, the policy is close to an open-loop forcing, whereas open-loop policies display a lesser efficiency for a reduced number of actuators, where the remaining load unsteadiness can be damped by an in-phase control action.

\begin{table}
    \centering
    \begin{tabular}{ccccccc}
        Nb. act. & $\alpha$ & $C_l$ & $C_d$ & $S(C_l)$ & $S(C_d)$ & $S(C_l)+S(C_d)$ \\
\hline\multirow{3}{*}{$10$} &  $12$ & $\textcolor{black}{0.0\%}$ & $\textcolor{black}{0.0\%}$ & $\textcolor{black}{0.0\%}$ & $\textcolor{black}{0.0\%}$ & $\textcolor{black}{0.0\%}$ \\ 
&  $15$ & $\textcolor{green}{+2.1\%}$ & $\textcolor{red}{+1.5\%}$ & $\textcolor{red}{+91\%}$ & $\textcolor{red}{+6.8\%}$ & $\textcolor{red}{+70\%}$ \\ 
&  $20$ & $\textcolor{green}{+0.9\%}$ & $\textcolor{red}{+0.8\%}$ & $\textcolor{red}{+37\%}$ & $\textcolor{red}{+57\%}$ & $\textcolor{red}{+42\%}$ \\ 
\hline\multirow{3}{*}{$5$} &  $12$ & $\textcolor{green}{+4.3\%}$ & $\textcolor{red}{+2.6\%}$ & $\textcolor{red}{+154\%}$ & $\textcolor{red}{+132\%}$ & $\textcolor{red}{+151\%}$ \\ 
&  $15$ & $\textcolor{green}{+2.9\%}$ & $\textcolor{red}{+2.1\%}$ & $\textcolor{red}{+120\%}$ & $\textcolor{red}{+98\%}$ & $\textcolor{red}{+115\%}$ \\ 
&  $20$ & $\textcolor{green}{+1.8\%}$ & $\textcolor{red}{+1.7\%}$ & $\textcolor{red}{+47\%}$ & $\textcolor{red}{+43\%}$ & $\textcolor{red}{+46\%}$ \\ 
\hline\multirow{3}{*}{$2$} &  $12$ & $\textcolor{green}{+5.1\%}$ & $\textcolor{red}{+2.9\%}$ & $\textcolor{red}{+126\%}$ & $\textcolor{red}{+107\%}$ & $\textcolor{red}{+123\%}$ \\ 
&  $15$ & $\textcolor{green}{+4.6\%}$ & $\textcolor{red}{+3.4\%}$ & $\textcolor{red}{+89\%}$ & $\textcolor{red}{+86\%}$ & $\textcolor{red}{+88\%}$ \\ 
&  $20$ & $\textcolor{green}{+1.8\%}$ & $\textcolor{red}{+1.7\%}$ & $\textcolor{red}{+12\%}$ & $\textcolor{red}{+24\%}$ & $\textcolor{red}{+14\%}$ \\
    \end{tabular}
    \caption{Comparison of the final (in the ``stabilised phase") performances in closed-loop and open-loop conditions. Performance variations are measured as relative variation with respect to the closed-loop performance.  \textcolor{green}{Green} coloured figures indicate that open-loop control performs better on the metric than closed-loop whereas \textcolor{red}{red} coloured ones state the opposite. Ensemble averages are computed on batches of 10 test runs.}
    \label{tab:NACA_ol_vs_cl}
\end{table}

To conclude, the relative improvement obtained by using a closed-loop strategy over an open-loop one is significant here. But regarding absolute gains, they are more marginal. Therefore, it may be feasible to get satisfactory control performances on this case through a parametric open-loop control study (with constant or variable suction), where the number, action amplitude and sign, and actuator position would be varied. Yet, even for this simple flow configuration, such an exhaustive study would face the ``curse of dimensionality" inherited from the combinatorial search of optimal layout (with, for instance, 252 different layouts having 5 active action components). Consequently, if carried out blindly with respect to the flow physics, this search would end up being significantly more expensive than the present algorithm, even for this relatively easy-to-control flow.

\section{Conclusion}
This study aims at proposing and testing a method of actuator selection, built on-top of a pre-trained on-policy reinforcement learning agent. This method relies on a sequential elimination of actuators based on a "what-if" analysis on the removal of each action component and selecting the component having the smallest estimated impact on performance. This process is repeated (after a "recovery" period) until the prescribed number of actuators is reached.

This method has first been applied to the one-dimensional periodic Kuramoto-Sivashinsky (KS) equation, starting from eight evenly distributed actuators. Their number has been successively reduced to one. The proposed elimination algorithm yields actuator layouts and sparse policies performing reasonably well compared to the maximum foreseeable performance, highlighting the ability of this method to provide a relevant approximation of the Pareto optimum of the performance with respect to the number of actuators at a lower cost than the exhaustive combinatorial study. The reasons for not reaching optimal performances have been investigated in details. This study has shown that, despite being a computationally cheap test-case, the KS equation posed numerous challenges to the method that are to be addressed in forthcoming developments, highlighting interesting research directions for the future. One may be the development of more complex approaches than the one-by-one elimination strategy that is proposed here. This strategy is indeed problematic when estimations are close to one another and where elimination choices lead to very different layouts, policies, and performances at the end of the selection process. 

The "what-if" analysis paradigm may also show limitations for cases where actuators can compensate for other eliminated ones since, the method does not account for the adaptation of the policy following the elimination, as illustrated by the KS equation test-case. One may imagine that the policy adaptation may retrospectively change the actuators' ranking in some cases. The strategy would be more accurate if it could account for it, or if it could consider the following elimination choices, mimicking a chess player thinking two or three moves ahead. In the present framework, this is unfortunately not tractable, since it would require around $n_{act}$ copies of the agent that would need to be trained in the same way the reference agent is. 
Therefore, the design of viable alternative strategies is a complex question that may be very interesting to explore in the future.

Then, a stalled bi-dimensional flow around a NACA 0012 airfoil has been considered as a control test case more representative of fluid mechanics control applications, both in terms of computational costs and dimensionality (even though it remains far from actual industrial flow configurations). The pre-trained controller showed first a nearly perfect stabilisation of the flow for the studied angles of attack. The method again succeeded at eliminating actuators while keeping a proficient level of stabilisation performance. Despite being controllable efficiently using an appropriate constant open-loop control, this test case underlines again the relevance of this method to finding an action-parsimonious control policy (in terms of number, position, and amplitude). 


Yet, as is, the computational cost of the present algorithm makes it \red{very expensive} for actual industrial flow configurations especially if a very large number of actuators (before elimination) is considered. Note that, to our knowledge, this is unfortunately the case for all state-of-the-art RL approaches for flow control.
The overhead cost of the present algorithm stems from the two following reasons. The first concerns the one-by-one elimination design choice, that forces a sequential elimination and imposes a minimum number of training epochs in order to reach the prescribed number of actuators. On the bright side, this strategy is useful when one does not know the final required number of actuators. Thus, our method can estimate the best layouts for each number of actuators and the corresponding performances. But again, one may want to derive a parallel elimination method that directly reaches the prescribed number of actuators.  Such a method would have to account for the policy adaptation issue previously highlighted. The second factor of extra cost is due to the need to converge the different value estimates for each elimination candidate using data from dedicated roll-outs. Measures have already been implemented to mitigate this cost, e.g. already eliminated components are systematically skipped. Yet, this cost could be divided by a factor of $n_{act}$ if the estimates used for ranking actuators could be converged using data from "standard" roll-outs, in an "on-the-fly" fashion. Therefore, future work may search for accurate ways to estimate the value function of sparse policies without dedicated roll-outs. This is a challenging question, and several attempts in this direction for the present study have not yield satisfactory results yet (not shown here). \red{At last, the mirroring issue of completing a ``deficient" or partial actuator layout could also be considered for future work. But this may require much more complex methods and paradigms since estimating the impact of an extra forcing action with its own optimised dynamics would require much more knowledge on the environment's dynamics. The issue of redistributing an initially ``unbalanced" layout also falls into this category.}

All these identified issues are as many ways to further improve a method tackling the crucial topic of optimal actuator layout and control behaviour, especially in the context of a gradual scale-up in the complexity of cases, towards more realistic cases and "industry-ready" control methods.

\section*{Acknowledgements}
This work is funded by the French Agency for Innovation and Defence (AID) via a PhD scholarship. Their support is gratefully acknowledged.

\section*{Declaration of Interests}
The authors report no conflict of interest.
\section*{Appendices}
\renewcommand*{\MyPath}{Z_Appendices}

\subsection{Hyper-parameters}
Table \ref{tab:params} presents the main numerical parameters of both the simulated case and the learning algorithm, with the notations used in the article (if introduced).
\begin{table}
\centering
\begin{tabular}{cccc}
 Parameter & Symbol & Value & Comment/Reference \\
 \hline
\multicolumn{4}{c}{\textbf{PPO-CMA hyper-parameters}}\\
Actor architecture & $\pi$ & $(512\times512)$ & 2 fully connected layers\\
Critic architecture & $V$ & $(512\times512)$ & 2 fully connected layers\\
Return discount factor & $\gamma$& $0.99$ & - \\
GAE control parameter & $\lambda_{GAE}$ & $0.97$ & Standard value\\
Optimiser & - & ADAM & \citet{Kingma2014}\\
History buffer depth & $H$ & $10$ epochs & - \\
\hline
\multicolumn{4}{c}{\textbf{AS-PPO-CMA hyper-parameters}}\\
Elimination duration & -& $200$ epochs & Fixed manually (phase (iii)) \\
\hline
\multicolumn{4}{c}{\textbf{KS hyper-parameters}}\\
Roll-out (episode) length & - & 250 & - \\
Roll-outs per epoch & - & 4 & - \\
Parallelized environments & - & 4 & $4\times4\times500 = 8000$ samples/epoch \\
Phase stability threshold & - & $5\%$ & Relative stability\\
CPU time per epoch & - & $\approx200$ s & Estimation \\
Total CPU time & - & $\approx78\times10^3$ CPUh. & for the whole batch of $200$ test cases\\
\hline
\multicolumn{4}{c}{\textbf{NACA0012 hyper-parameters}}\\
Roll-out (episode) length & - & 250 & - \\
Roll-outs per epoch & - & 1 & - \\
Parallelized environments & - & 5 & $5\times1\times250=1250$ samples/epochs\\
Phase stability threshold & - & $2\%$ & Relative stability\\
CPU time per epoch & - & $\approx3.6$ h & Estimation \\
Total CPU time & - & $\approx645\times10^3$ CPUh. & for the whole batch of $30$ test cases\\
\end{tabular}
\caption{\label{tab:params} Additional numerical parameters}
\end{table}
\subsection{Elimination speed}
\begin{figure}
    \centering
    \includegraphics[width=.6\textwidth]{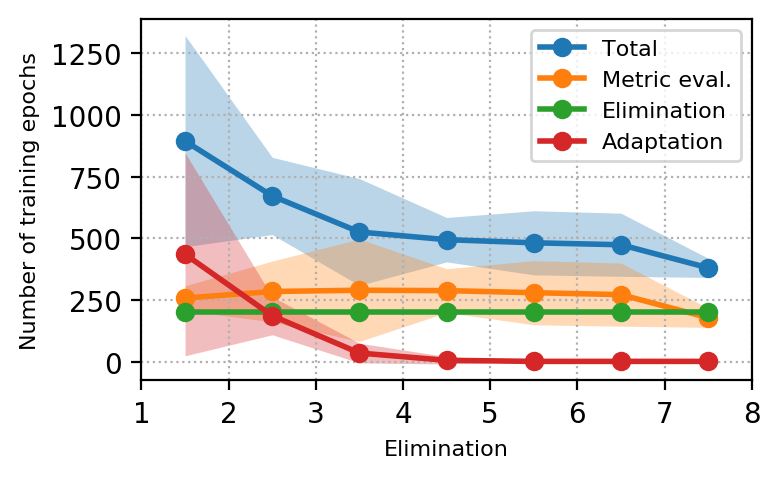}
    \caption{Duration of the elimination phases with respect to the eliminated action component. The ensemble average (solid line) is computed across the 200-test cases batch. The shaded area illustrates the corresponding ensemble standard deviation.}
    \label{fig:KS_elim_speed}
\end{figure}

\red{Figure \ref{fig:KS_elim_speed} illustrates the average cost of elimination depending on the eliminated action component. This cost is the number of training epochs required to run phases 2 (metric evaluation), 3 (elimination) and 1 (policy adaptation), in that order, for a given elimination. One can notice a rather stable overhead of around $500$ epochs from elimination $8\rightarrow7$ to $4\rightarrow3$, then an inflation of this cost for the last two ones. This increase is mainly due to the policy adaptation phase that takes longer with few actuators.}
\red{This behaviour can be correlated with the slower convergence of the learning curves for layouts having $1$ and $2$ actuators compared to the others, in the systematic study (not shown here).}

\subsection{Performances with 1 actuator on the KS equation}\label{app:KS_1act}
\begin{figure}
    \centering
    \includegraphics[width=\textwidth]{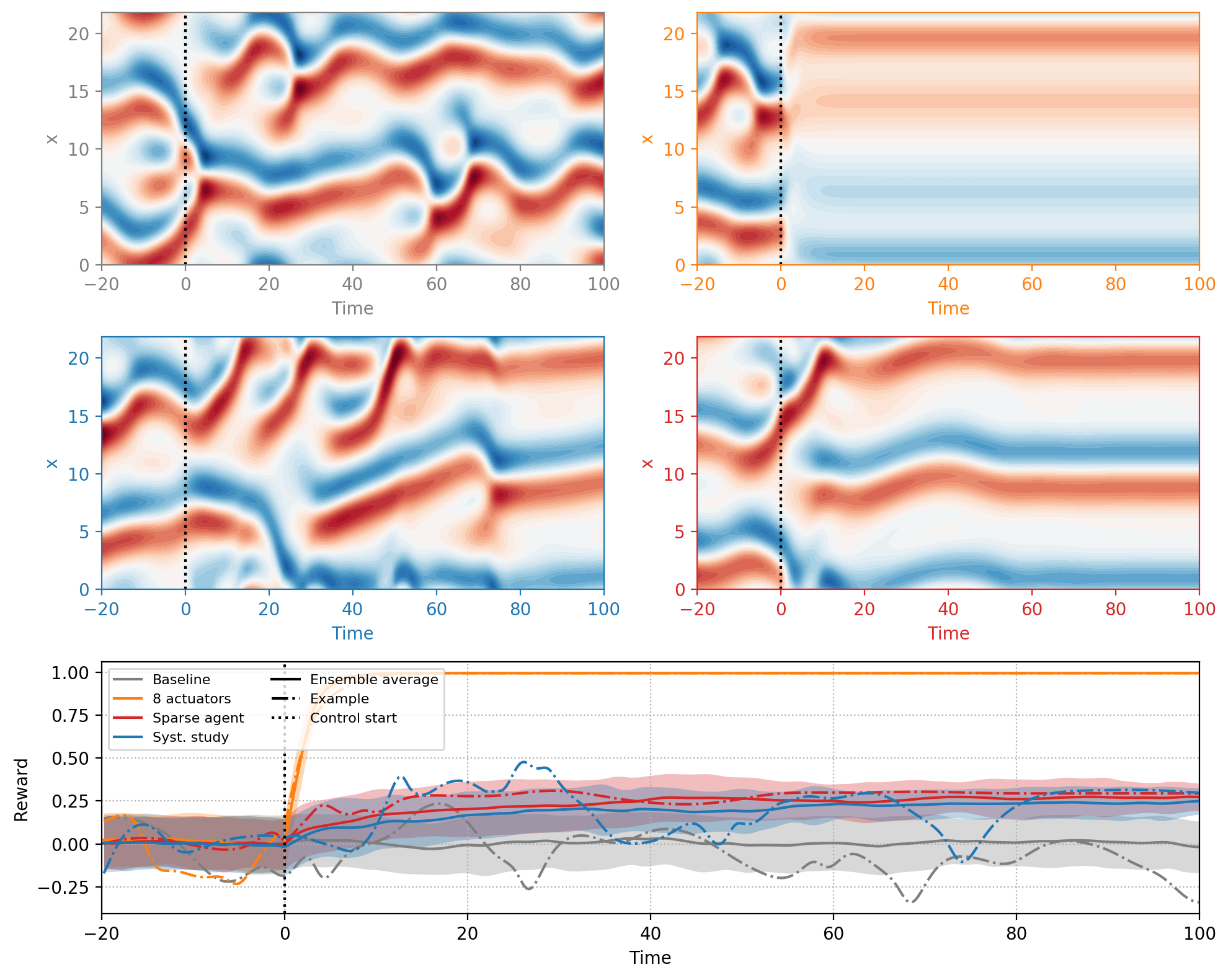}
    \caption{(bottom) Comparison of the performances of both 1-actuator agents (having only actuator 0 activated) obtained via the systematic study (blue line) and the elimination process (red line) with the non-controlled baseline (grey line) and an 8-actuator agent (orange line). Solid lines represent ensemble averages computed over batches over 100 runs, shaded areas account of the standard deviation across the batch, while dash-dotted lines depict the example run reported on the upper graphs. (upper graphs) Evolution of the state on example runs for each agent.}
    \label{fig:KS_1act_perfo}
\end{figure}

With only one actuator, the obtained layouts slightly over-perform the systematic study best individual as shown in figure~\ref{fig:KS_1act_perfo}. \red{As discussed, this moderate over-performance of the agents obtained by the elimination process compared to the ones converged in the systematic study may be attributed to a better exploration of the state-action space during training. Comparing both example runs (blue-framed and red-framed graphs), one can notice than both agents tend to stabilise the state around the same target. While still being sub-optimal, this stationary state is close to the second fixed point $E_2$ (if $E_2$ were the control target, the reward would be around $0.72$) and is still better performing than the average non-controlled state. The gap in performance appears to be the consequence of a better stability of the state for the ``sparsified" agent.}

\subsection{Other results on the NACA test case}\label{app:NACA}

\begin{figure}
    \centering
    \includegraphics[width=\textwidth]{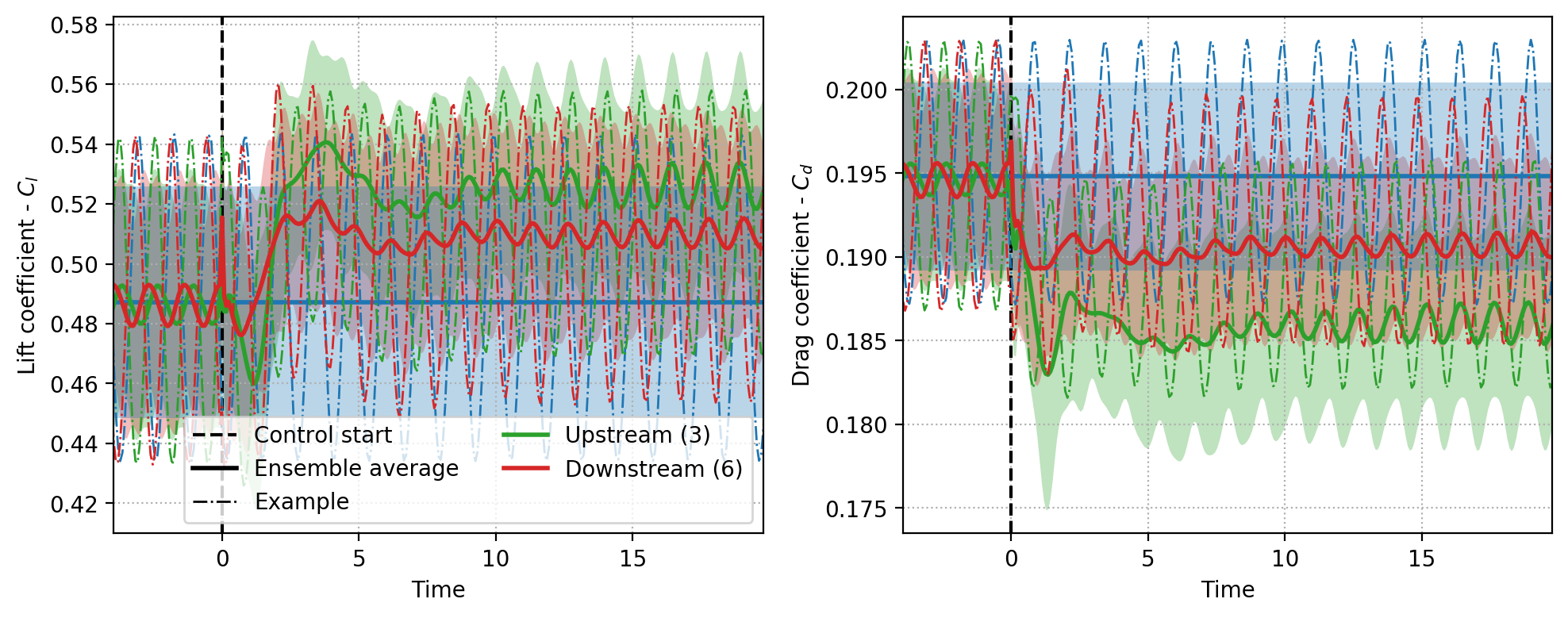}
    \caption{Comparison of lift and drag open-loop control performances using actuator number 3 (as prescribed by the sparsification method), upstream to the separation range or using actuator number 6, located nearby the separation range. Ensemble averages are computed on a batch of 40 test runs.}
    \label{fig:NACA_12deg_upst_vs_dnst}
\end{figure}

Figure \ref{fig:NACA_sparse_hist_comp} shows the observed actuator frequency selection with respect to the total number of active components and the angle of attack. This selection appears to be, as expected, correlated with the position of the uncontrolled flow separation range for $\alpha=15$° and $\alpha=20$°, whereas for $\alpha=12$°, the sparsification method keeps action components significantly upstream from the separation range. Figure \ref{fig:NACA_12deg_upst_vs_dnst} shows the difference in lift and drag performance for a constant open-loop forcing action implemented respectively on actuator 3 and 6. One can notice a significant advantage of implementing the forcing on the upstream actuator number 3 rather than actuator 6, thus justifying the elimination choice of the algorithm.

\begin{figure}
    \centering
    \includegraphics[width=.8\textwidth]{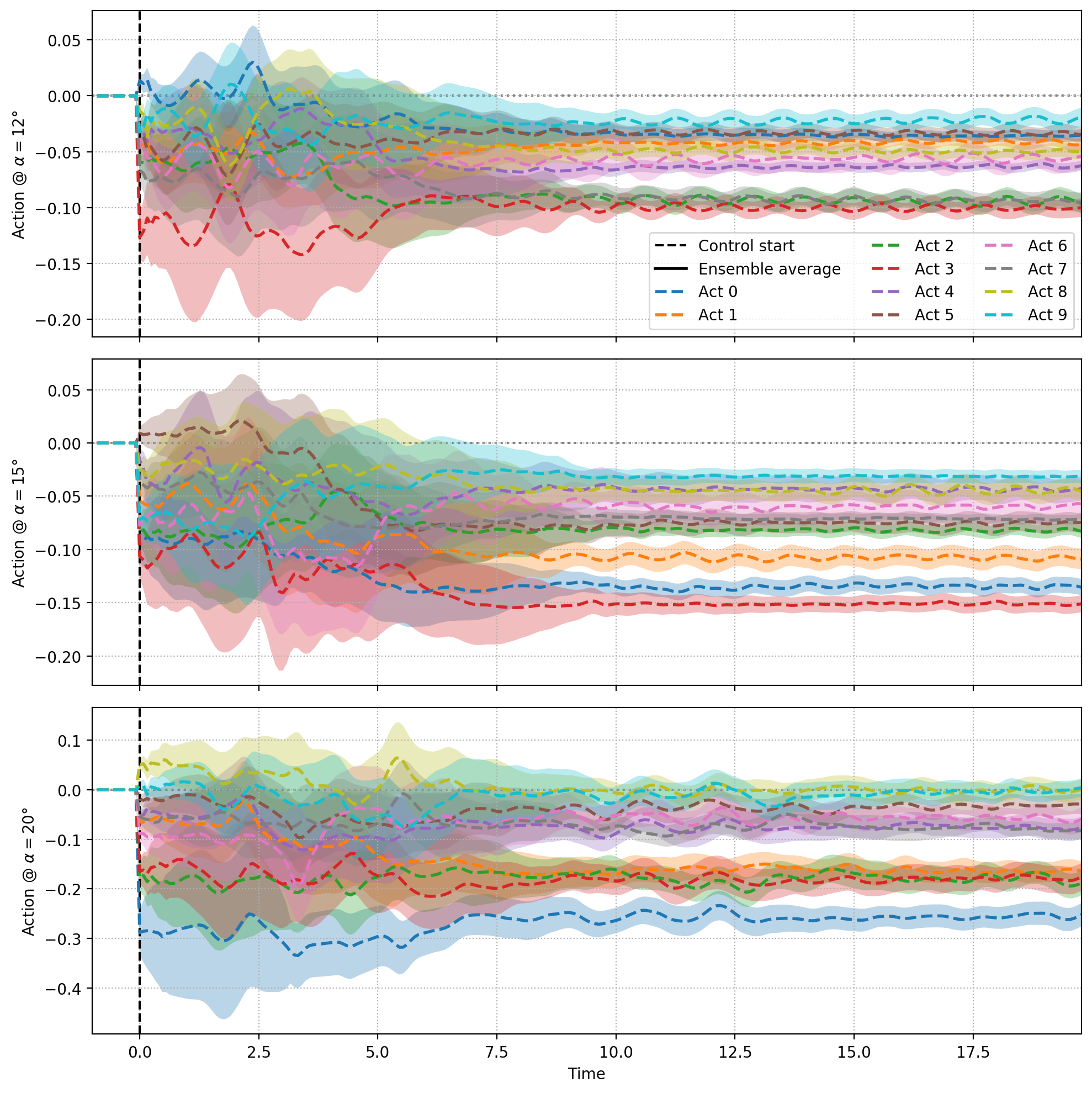}
    \caption{Ensemble average of control actions of the pre-trained policies for $\alpha=12$° (top), $\alpha=15$° (middle) and $\alpha=20$° (bottom). Averages are computed on batches of 50 test-runs. Shaded areas illustrate the standard deviation across the batch. Action component numbering follows the notation introduced by figure \ref{fig:NACA_domain}.}
    \label{fig:NACA_act_ampl}
\end{figure}

Figure \ref{fig:NACA_act_ampl} compares the ensemble average of the control actions on test-runs of the pre-trained agents (using all the available action components) for the different angles of attack.

\bibliographystyle{jfm}
\bibliography{biblioThese}

\end{document}